\documentclass[11pt]{amsart}

\pdfoutput=1

\usepackage[utf8]{inputenc}  

\usepackage[T1]{fontenc}  

\usepackage[text={400pt,660pt},centering]{geometry}

\usepackage{esint,amssymb}
\usepackage{graphicx}
\usepackage{MnSymbol}
\usepackage{mathtools}
\usepackage[colorlinks=true, pdfstartview=FitV, linkcolor=blue, citecolor=blue, urlcolor=blue,pagebackref=false]{hyperref}
\usepackage{microtype}

\usepackage{bm}
\usepackage{dsfont}

\newtheorem{rmk}{Remarque}

\theoremstyle{remark}

\theoremstyle{definition}

\numberwithin{equation}{section}
\numberwithin{proposition}{section}
\numberwithin{figure}{section}
\numberwithin{table}{section}

\newcommand{\Z}{\mathbb{Z}}

\newcommand{\R}{\mathbb{R}}

\renewcommand{\le}{\leqslant}
\renewcommand{\ge}{\geqslant}
\renewcommand{\leq}{\leqslant}
\renewcommand{\geq}{\geqslant}

\DeclareMathOperator{\cotan}{cotan}

\renewcommand{\tilde}{\widetilde}

\newcommand{\jr}{\tau}
\newcommand{\hr}{\omega}
\newcommand{\obl}{\varepsilon}
\newcommand{\lat}{\phi}
\newcommand{\ssi}{si et seulment si}
\newcommand{\Rdeux}{R^{(2)}}
\newcommand{\Rtrois}{R^{(3)}}
\newcommand{\bigmatrice}{\mathfrak{M}}

\begin{document}

\author[A. Bordas]{A. Bordas}
\address[A. Bordas]{Lycée Bellepierre, Saint Denis, Réunion, France}
\email{alexandre.bordas@ac-reunion.fr}

\date{\today}

\title[Day length, solar received energy]{Computation of day length, amount of energy received over a day and over a year in function of latitude, date and obliquity.}

\begin{abstract}
	In these notes, we do the computation of the formula giving the expression of solar energy received by an horizontal section of 1 square meter, relatively to the latitude, obliquity and current day (the quantity is also proportionnal to the solar flux). 
	We also obtain the formula of energy received during one day, of day length, of the possibility, according to the latitude, to observe the sun at zenith one day in year (and in this case, the formula of the day it happens), the possibility to observe a oplar night/day, and the time of beginning of these phenomenons according to the latitude. 
	
Section \ref{s.notations} introduce the $4$ different angles which are obliquity, latitude, day and hour. We also introduce the corresponding rotations.

	In \ref{s.positions_relatives}, we give the position of the sun in the sky.
	That make able to compute the day length in \ref{s.duree_jour}
	Following sections are pairwie independent, except \ref{s.energie_annee} which relies on \ref{s.energiejour}.
	In\ref{s.nuit_jour_polaires}, we compute for which latitudes it is possible to observe or not a polar day or a polar night, and give the time of beginning and en of these lasts.
	In \ref{s.energiejour}, we intergate between sunrise and sunset the received solar power, which give us the quantity of energy received during one day.
	In \ref{s.energie_annee}, we integrate the following quantity in order to obtain the quantity of energy received per year.
	In \ref{s.zenith}, we prove that only latitudes located between tropics can observe sun at zenith, and give the two instants it happens in a year.
	In \ref{s.lever_coucher_soleil} we compute the direction of sunrise and sunset, given atitude and day.
\end{abstract}

\maketitle

\section{Notations}\label{s.notations}

We are placing in the orthonoraml base in which origin is the center of earth. $x$ axis is directed to the sun, $y$ axis belong to erath revolution plan and is directed to west, $z$ axis is directed to north pole.

In every point $M$ of the sphere, a square of surface 1 square meter receives a quantity of energy which is proportional to the scalar product of the unit orthonromal of this plan (which is precisely $\overrightarrow{OM}$) and the unit vector directed to the sun (which is $e_x=\begin{pmatrix}).
1\\0\\0
\end{pmatrix}$.
\begin{itemize}
	\item We denote by $\obl $ the axial tilte of earth (actually, that quantity is not constant, but oscillates around its mean value) $\obl \sim 23^\circ$. 
	Rotation of $Oy$ axis of angle $\obl$ has the following matrix
	\[\Rdeux_{\obl}=\begin{pmatrix}
	\cos(\obl) & 0 & -\sin(\obl)\\
	0&1&0\\
	\sin(\obl) & 0 & \cos(\obl)
	\end{pmatrix}.\]
	 \begingroup%
	\makeatletter%
	\providecommand\color[2][]{%
		\errmessage{(Inkscape) Color is used for the text in Inkscape, but the package 'color.sty' is not loaded}%
		\renewcommand\color[2][]{}%
	}%
	\providecommand\transparent[1]{%
		\errmessage{(Inkscape) Transparency is used (non-zero) for the text in Inkscape, but the package 'transparent.sty' is not loaded}%
		\renewcommand\transparent[1]{}%
	}%
	\providecommand\rotatebox[2]{#2}%
	\newcommand*\fsize{\dimexpr\f@size pt\relax}%
	\newcommand*\lineheight[1]{\fontsize{\fsize}{#1\fsize}\selectfont}%
	\ifx\svgwidth\undefined%
	\setlength{\unitlength}{285.5950622bp}%
	\ifx\svgscale\undefined%
	\relax%
	\else%
	\setlength{\unitlength}{\unitlength * \real{\svgscale}}%
	\fi%
	\else%
	\setlength{\unitlength}{\svgwidth}%
	\fi%
	\global\let\svgwidth\undefined%
	\global\let\svgscale\undefined%
	\makeatother%
	\begin{picture}(1,0.81405319)%
	\lineheight{1}%
	\setlength\tabcolsep{0pt}%
	\put(0,0){\includegraphics[width=\unitlength,page=1]{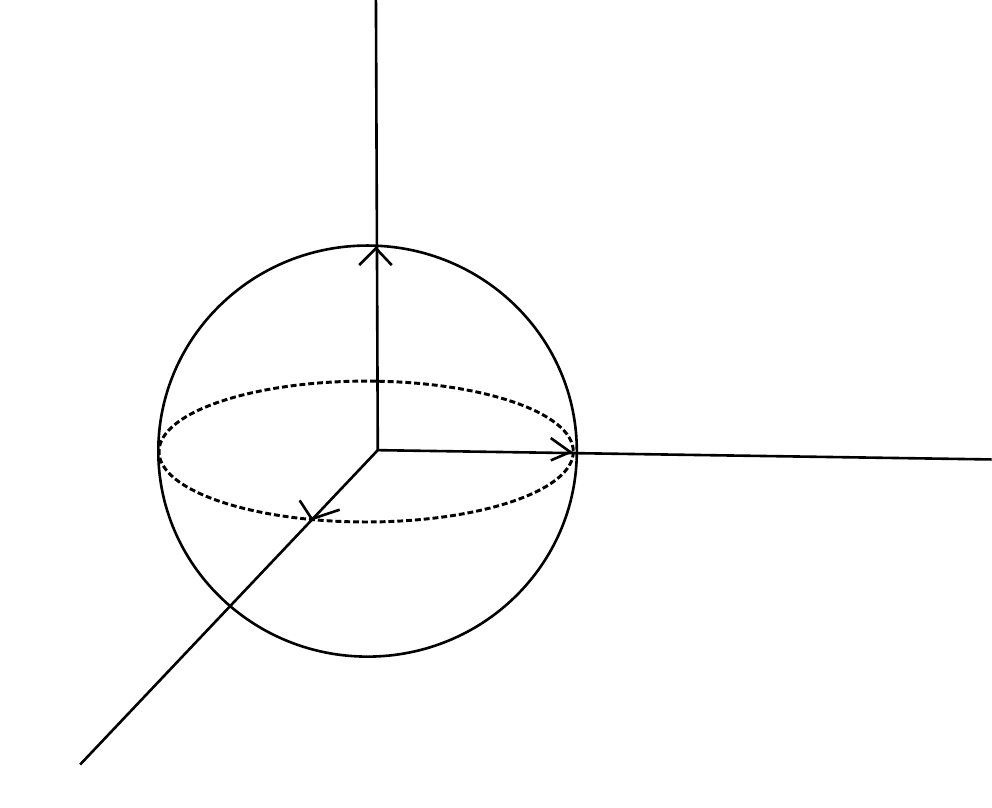}}%
	\put(0.32949814,0.59842698){\color[rgb]{0,0,0}\makebox(0,0)[lt]{\lineheight{1.25}\smash{\begin{tabular}[t]{l}z\end{tabular}}}}%
	\put(0.78264452,0.31719293){\color[rgb]{0,0,0}\makebox(0,0)[lt]{\lineheight{1.25}\smash{\begin{tabular}[t]{l}y\end{tabular}}}}%
	\put(0.09866653,0.02809487){\color[rgb]{0,0,0}\makebox(0,0)[lt]{\lineheight{1.25}\smash{\begin{tabular}[t]{l}x\end{tabular}}}}%
	\put(0.27123851,0.33431628){\color[rgb]{0,0,0}\makebox(0,0)[lt]{\lineheight{1.25}\smash{\begin{tabular}[t]{l}$e_x$\end{tabular}}}}%
	\put(0.47889273,0.33620365){\color[rgb]{0,0,0}\makebox(0,0)[lt]{\lineheight{1.25}\smash{\begin{tabular}[t]{l}$e_y$\end{tabular}}}}%
	\put(0.31173242,0.52590442){\color[rgb]{0,0,0}\makebox(0,0)[lt]{\lineheight{1.25}\smash{\begin{tabular}[t]{l}$e_z$\end{tabular}}}}%
	\put(0.68297281,0.43279487){\color[rgb]{0,0,0}\makebox(0,0)[lt]{\lineheight{1.25}\smash{\begin{tabular}[t]{l} \end{tabular}}}}%
	\put(0,0){\includegraphics[width=\unitlength,page=2]{rotation_epsilon.pdf}}%
	\put(0.55100788,0.61367735){\color[rgb]{0,0,0}\makebox(0,0)[lt]{\lineheight{1.25}\smash{\begin{tabular}[t]{l}North pole\end{tabular}}}}%
	\put(0,0){\includegraphics[width=\unitlength,page=3]{rotation_epsilon.pdf}}%
	\put(0.71823407,0.4005664){\color[rgb]{0,0,0}\makebox(0,0)[lt]{\lineheight{1.25}\smash{\begin{tabular}[t]{l}$\obl$\end{tabular}}}}%
	\put(0,0){\includegraphics[width=\unitlength,page=4]{rotation_epsilon.pdf}}%
	\put(0.4023554,0.44526402){\color[rgb]{0,0,0}\makebox(0,0)[lt]{\lineheight{1.25}\smash{\begin{tabular}[t]{l}$\obl$\end{tabular}}}}%
	\put(0.36648107,0.32028993){\color[rgb]{0,0,0}\makebox(0,0)[lt]{\lineheight{1.25}\smash{\begin{tabular}[t]{l}$\obl$\end{tabular}}}}%
	\put(0.63997763,0.19696739){\color[rgb]{0,0,0}\makebox(0,0)[lt]{\lineheight{1.25}\smash{\begin{tabular}[t]{l}Equator\end{tabular}}}}%
	\put(0,0){\includegraphics[width=\unitlength,page=5]{rotation_epsilon.pdf}}%
	\put(0.23715523,0.40344727){\color[rgb]{0,0,0}\makebox(0,0)[lt]{\lineheight{1.25}\smash{\begin{tabular}[t]{l}$R^2_\obl e_x$\end{tabular}}}}%
	\put(0.40846636,0.49809692){\color[rgb]{0,0,0}\makebox(0,0)[lt]{\lineheight{1.25}\smash{\begin{tabular}[t]{l}$R^2_\obl e_z$\end{tabular}}}}%
	\put(0,0){\includegraphics[width=\unitlength,page=6]{rotation_epsilon.pdf}}%
	\put(0.54422862,0.66422365){\color[rgb]{0,0,0}\makebox(0,0)[lt]{\lineheight{1.25}\smash{\begin{tabular}[t]{l}Earth's axis of rotation\end{tabular}}}}%
	\put(0,0){\includegraphics[width=\unitlength,page=7]{rotation_epsilon.pdf}}%
	\end{picture}%
	\endgroup%
	\item We denote by $\lat \in [0,\pi/2]$ the latitude of the considered point (if $\lat\in[-\pi/2,0]$, the observed phenomenons are the same as fo $-\lat$, with a $6$ months time shift), i.e.\[M_\lat := \begin{pmatrix}
	\cos(\lat)\\
	0\\
	\sin(\lat)
	\end{pmatrix}=\Rdeux_{\lat}.\begin{pmatrix}
	1\\0\\0
	\end{pmatrix}. \]
	 \begingroup%
	\makeatletter%
	\providecommand\color[2][]{%
		\errmessage{(Inkscape) Color is used for the text in Inkscape, but the package 'color.sty' is not loaded}%
		\renewcommand\color[2][]{}%
	}%
	\providecommand\transparent[1]{%
		\errmessage{(Inkscape) Transparency is used (non-zero) for the text in Inkscape, but the package 'transparent.sty' is not loaded}%
		\renewcommand\transparent[1]{}%
	}%
	\providecommand\rotatebox[2]{#2}%
	\newcommand*\fsize{\dimexpr\f@size pt\relax}%
	\newcommand*\lineheight[1]{\fontsize{\fsize}{#1\fsize}\selectfont}%
	\ifx\svgwidth\undefined%
	\setlength{\unitlength}{285.5950622bp}%
	\ifx\svgscale\undefined%
	\relax%
	\else%
	\setlength{\unitlength}{\unitlength * \real{\svgscale}}%
	\fi%
	\else%
	\setlength{\unitlength}{\svgwidth}%
	\fi%
	\global\let\svgwidth\undefined%
	\global\let\svgscale\undefined%
	\makeatother%
	\begin{picture}(1,0.81405319)%
	\lineheight{1}%
	\setlength\tabcolsep{0pt}%
	\put(0,0){\includegraphics[width=\unitlength,page=1]{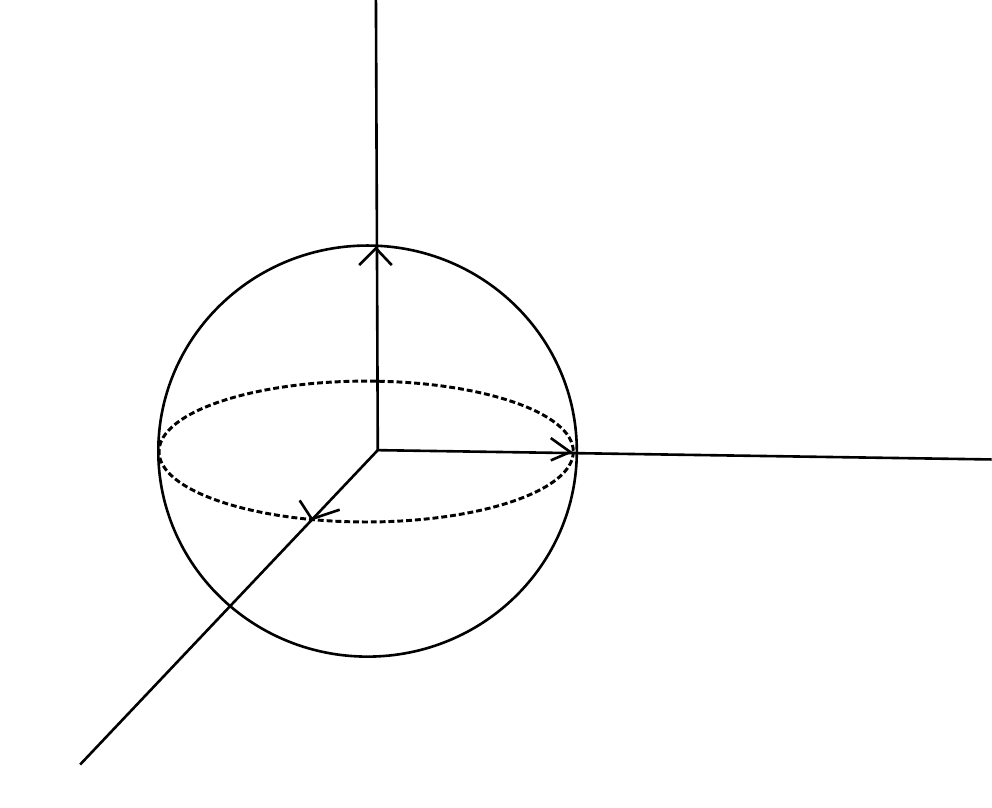}}%
	\put(0.32949814,0.59842698){\color[rgb]{0,0,0}\makebox(0,0)[lt]{\lineheight{1.25}\smash{\begin{tabular}[t]{l}z\end{tabular}}}}%
	\put(0.78264452,0.31719293){\color[rgb]{0,0,0}\makebox(0,0)[lt]{\lineheight{1.25}\smash{\begin{tabular}[t]{l}y\end{tabular}}}}%
	\put(0.09866653,0.02809487){\color[rgb]{0,0,0}\makebox(0,0)[lt]{\lineheight{1.25}\smash{\begin{tabular}[t]{l}x\end{tabular}}}}%
	\put(0.27123851,0.33431628){\color[rgb]{0,0,0}\makebox(0,0)[lt]{\lineheight{1.25}\smash{\begin{tabular}[t]{l}$e_x$\end{tabular}}}}%
	\put(0.47889273,0.33620365){\color[rgb]{0,0,0}\makebox(0,0)[lt]{\lineheight{1.25}\smash{\begin{tabular}[t]{l}$e_y$\end{tabular}}}}%
	\put(0.31173242,0.52590442){\color[rgb]{0,0,0}\makebox(0,0)[lt]{\lineheight{1.25}\smash{\begin{tabular}[t]{l}$e_z$\end{tabular}}}}%
	\put(0.68297281,0.43279487){\color[rgb]{0,0,0}\makebox(0,0)[lt]{\lineheight{1.25}\smash{\begin{tabular}[t]{l} \end{tabular}}}}%
	\put(0,0){\includegraphics[width=\unitlength,page=2]{rotation_phi.pdf}}%
	\put(0.17826791,0.57304902){\color[rgb]{0,0,0}\makebox(0,0)[lt]{\lineheight{1.25}\smash{\begin{tabular}[t]{l}North pole\end{tabular}}}}%
	\put(0,0){\includegraphics[width=\unitlength,page=3]{rotation_phi.pdf}}%
	\put(0.71823407,0.4005664){\color[rgb]{0,0,0}\makebox(0,0)[lt]{\lineheight{1.25}\smash{\begin{tabular}[t]{l}$\phi$\end{tabular}}}}%
	\put(0,0){\includegraphics[width=\unitlength,page=4]{rotation_phi.pdf}}%
	\put(0.4023554,0.44526402){\color[rgb]{0,0,0}\makebox(0,0)[lt]{\lineheight{1.25}\smash{\begin{tabular}[t]{l}$\phi$\end{tabular}}}}%
	\put(0.36648107,0.32028993){\color[rgb]{0,0,0}\makebox(0,0)[lt]{\lineheight{1.25}\smash{\begin{tabular}[t]{l}$\phi$\end{tabular}}}}%
	\put(0,0){\includegraphics[width=\unitlength,page=5]{rotation_phi.pdf}}%
	\put(0.63666172,0.59090175){\color[rgb]{0,0,0}\makebox(0,0)[lt]{\lineheight{1.25}\smash{\begin{tabular}[t]{l}Latitude $\phi$\end{tabular}}}}%
	\put(0.63997763,0.19696739){\color[rgb]{0,0,0}\makebox(0,0)[lt]{\lineheight{1.25}\smash{\begin{tabular}[t]{l}Equator\end{tabular}}}}%
	\end{picture}%
	\endgroup%
	\item 	We denote by $\jr\in [0,2\pi]$ the angle which corresponds to the number of days since the last winter solstice (around $21^{st}  of december$).
	\[\jr=2\pi \times\frac{Number~of~days~since~winter~solstice}{365.25} ,\]
	or expressed in degrees : $\jr_{degree}=\frac{180}{\pi} t = 360 \times\frac{Number~of~days~since~winter~solstice}{365.25}.$ 
	Rotation of $Oz$ axis of angle $\jr$ admits the following matrix
	\[
	\Rtrois_{-\jr}=
	\begin{pmatrix}
	\cos(\jr) & -\sin(\jr) & 0 \\
	\sin(\jr) & \cos(\jr)  & 0 \\
	0&0&1
	\end{pmatrix}.
	\]
	
	 \begingroup%
	\makeatletter%
	\providecommand\color[2][]{%
		\errmessage{(Inkscape) Color is used for the text in Inkscape, but the package 'color.sty' is not loaded}%
		\renewcommand\color[2][]{}%
	}%
	\providecommand\transparent[1]{%
		\errmessage{(Inkscape) Transparency is used (non-zero) for the text in Inkscape, but the package 'transparent.sty' is not loaded}%
		\renewcommand\transparent[1]{}%
	}%
	\providecommand\rotatebox[2]{#2}%
	\newcommand*\fsize{\dimexpr\f@size pt\relax}%
	\newcommand*\lineheight[1]{\fontsize{\fsize}{#1\fsize}\selectfont}%
	\ifx\svgwidth\undefined%
	\setlength{\unitlength}{262.77801437bp}%
	\ifx\svgscale\undefined%
	\relax%
	\else%
	\setlength{\unitlength}{\unitlength * \real{\svgscale}}%
	\fi%
	\else%
	\setlength{\unitlength}{\svgwidth}%
	\fi%
	\global\let\svgwidth\undefined%
	\global\let\svgscale\undefined%
	\makeatother%
	\begin{picture}(1,0.85919776)%
	\lineheight{1}%
	\setlength\tabcolsep{0pt}%
	\put(0,0){\includegraphics[width=\unitlength,page=1]{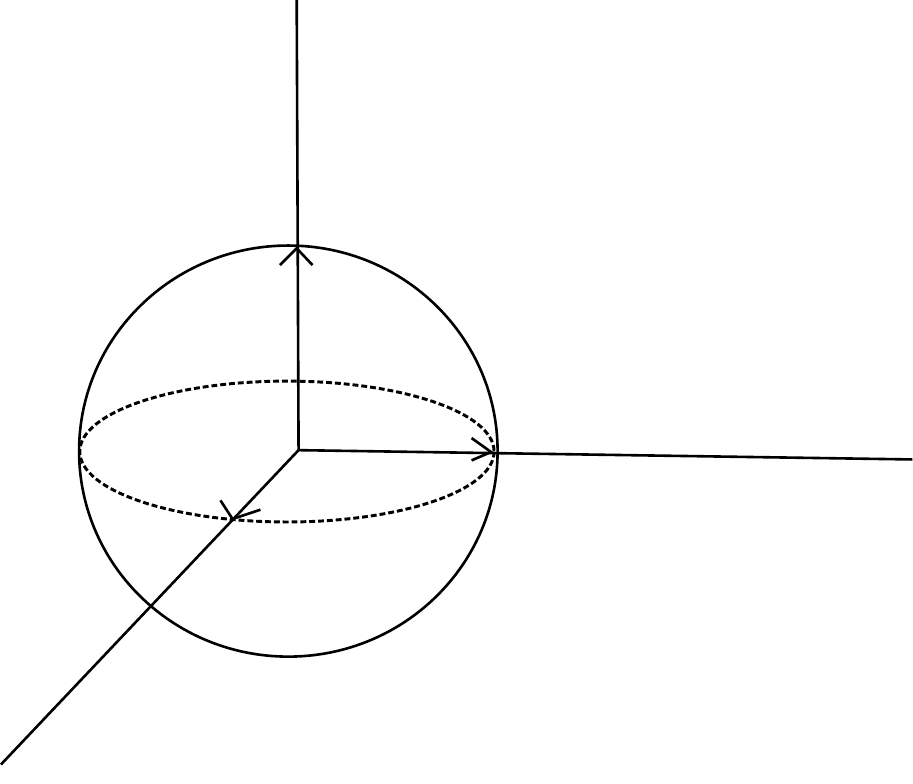}}%
	\put(0.27127833,0.62484867){\color[rgb]{0,0,0}\makebox(0,0)[lt]{\lineheight{1.25}\smash{\begin{tabular}[t]{l}z\end{tabular}}}}%
	\put(0.76377151,0.31919502){\color[rgb]{0,0,0}\makebox(0,0)[lt]{\lineheight{1.25}\smash{\begin{tabular}[t]{l}y\end{tabular}}}}%
	\put(0.02040357,0.00499452){\color[rgb]{0,0,0}\makebox(0,0)[lt]{\lineheight{1.25}\smash{\begin{tabular}[t]{l}x\end{tabular}}}}%
	\put(0.20796,0.3378052){\color[rgb]{0,0,0}\makebox(0,0)[lt]{\lineheight{1.25}\smash{\begin{tabular}[t]{l}$e_x$\end{tabular}}}}%
	\put(0.43364488,0.33985644){\color[rgb]{0,0,0}\makebox(0,0)[lt]{\lineheight{1.25}\smash{\begin{tabular}[t]{l}$e_y$\end{tabular}}}}%
	\put(0.25197001,0.54602897){\color[rgb]{0,0,0}\makebox(0,0)[lt]{\lineheight{1.25}\smash{\begin{tabular}[t]{l}$e_z$\end{tabular}}}}%
	\put(0.65544528,0.4448347){\color[rgb]{0,0,0}\makebox(0,0)[lt]{\lineheight{1.25}\smash{\begin{tabular}[t]{l} \end{tabular}}}}%
	\put(0.35710338,0.51580694){\color[rgb]{0,0,0}\makebox(0,0)[lt]{\lineheight{1.25}\smash{\begin{tabular}[t]{l}$R^2_\epsilon e_z$\end{tabular}}}}%
	\put(0.60871927,0.65376247){\color[rgb]{0,0,0}\rotatebox{29.49902531}{\makebox(0,0)[lt]{\lineheight{1.25}\smash{\begin{tabular}[t]{l}Earth's orbit\end{tabular}}}}}%
	\put(0,0){\includegraphics[width=\unitlength,page=2]{rotation_jour.pdf}}%
	\put(0.73672937,0.16237978){\color[rgb]{0,0,0}\makebox(0,0)[lt]{\lineheight{1.25}\smash{\begin{tabular}[t]{l}$-e_x$\\\end{tabular}}}}%
	\put(0.62787996,0.1962668){\color[rgb]{0,0,0}\makebox(0,0)[lt]{\lineheight{1.25}\smash{\begin{tabular}[t]{l}$\jr$\end{tabular}}}}%
	\put(0.5508637,0.14389586){\color[rgb]{0.9372549,0.88235294,0.05882353}\makebox(0,0)[lt]{\lineheight{1.25}\smash{\begin{tabular}[t]{l}$-R^3_\jr  e_x$\end{tabular}}}}%
	\put(0,0){\includegraphics[width=\unitlength,page=3]{rotation_jour.pdf}}%
	\end{picture}%
	\endgroup%
	
	\item We denote by $\hr$ the angle corresponding to the hour. 
	Rotation of  $Oz$ axis, of angle $\hr$ admits the following matrix
	\[
	\Rtrois_\hr=
	\begin{pmatrix}
	\cos(\hr) & -\sin(\hr) & 0 \\
	\sin(\hr) & \cos(\hr)  & 0 \\
	0&0&1
	\end{pmatrix}.
	\]
	 \begingroup%
	\makeatletter%
	\providecommand\color[2][]{%
		\errmessage{(Inkscape) Color is used for the text in Inkscape, but the package 'color.sty' is not loaded}%
		\renewcommand\color[2][]{}%
	}%
	\providecommand\transparent[1]{%
		\errmessage{(Inkscape) Transparency is used (non-zero) for the text in Inkscape, but the package 'transparent.sty' is not loaded}%
		\renewcommand\transparent[1]{}%
	}%
	\providecommand\rotatebox[2]{#2}%
	\newcommand*\fsize{\dimexpr\f@size pt\relax}%
	\newcommand*\lineheight[1]{\fontsize{\fsize}{#1\fsize}\selectfont}%
	\ifx\svgwidth\undefined%
	\setlength{\unitlength}{233.67281936bp}%
	\ifx\svgscale\undefined%
	\relax%
	\else%
	\setlength{\unitlength}{\unitlength * \real{\svgscale}}%
	\fi%
	\else%
	\setlength{\unitlength}{\svgwidth}%
	\fi%
	\global\let\svgwidth\undefined%
	\global\let\svgscale\undefined%
	\makeatother%
	\begin{picture}(1,0.77082962)%
	\lineheight{1}%
	\setlength\tabcolsep{0pt}%
	\put(0,0){\includegraphics[width=\unitlength,page=1]{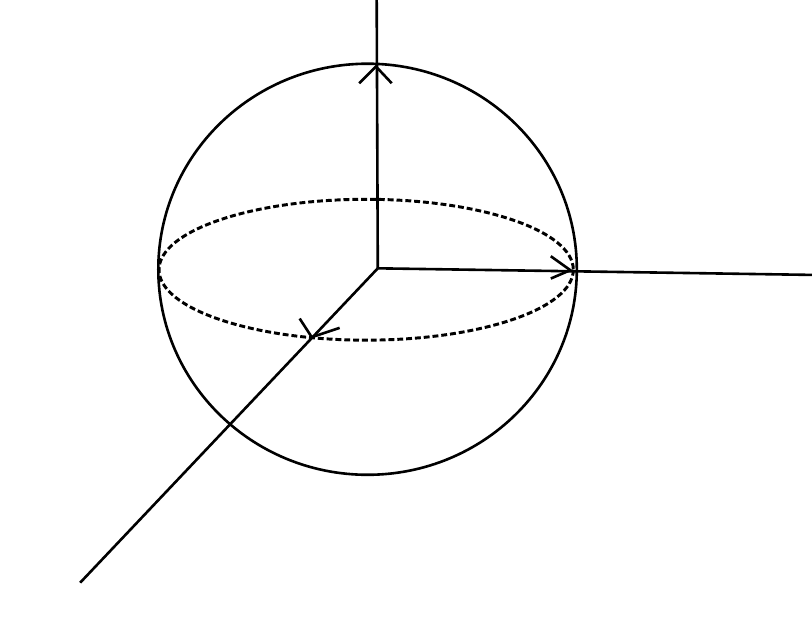}}%
	\put(0.40271284,0.73139777){\color[rgb]{0,0,0}\makebox(0,0)[lt]{\lineheight{1.25}\smash{\begin{tabular}[t]{l}z\end{tabular}}}}%
	\put(0.95654863,0.38767334){\color[rgb]{0,0,0}\makebox(0,0)[lt]{\lineheight{1.25}\smash{\begin{tabular}[t]{l}y\end{tabular}}}}%
	\put(0.12059029,0.03433751){\color[rgb]{0,0,0}\makebox(0,0)[lt]{\lineheight{1.25}\smash{\begin{tabular}[t]{l}x\end{tabular}}}}%
	\put(0,0){\includegraphics[width=\unitlength,page=2]{rotation_omega.pdf}}%
	\put(0.33150787,0.40860151){\color[rgb]{0,0,0}\makebox(0,0)[lt]{\lineheight{1.25}\smash{\begin{tabular}[t]{l}$e_x$\end{tabular}}}}%
	\put(0.53180672,0.40928713){\color[rgb]{0,0,0}\makebox(0,0)[lt]{\lineheight{1.25}\smash{\begin{tabular}[t]{l}$e_y$\end{tabular}}}}%
	\put(0.35653845,0.57595538){\color[rgb]{0,0,0}\makebox(0,0)[lt]{\lineheight{1.25}\smash{\begin{tabular}[t]{l}$e_z$\end{tabular}}}}%
	\put(0.50716134,0.286063){\color[rgb]{0.61568627,0,0}\makebox(0,0)[lt]{\lineheight{1.25}\smash{\begin{tabular}[t]{l}$R^3_\omega e_x$\end{tabular}}}}%
	\put(0.5784608,0.52861597){\color[rgb]{0.63529412,0,0}\makebox(0,0)[lt]{\lineheight{1.25}\smash{\begin{tabular}[t]{l}\(R^3_\omega e_y\)\end{tabular}}}}%
	\put(0.8347298,0.52896211){\color[rgb]{0,0,0}\makebox(0,0)[lt]{\lineheight{1.25}\smash{\begin{tabular}[t]{l} \end{tabular}}}}%
	\put(0,0){\includegraphics[width=\unitlength,page=3]{rotation_omega.pdf}}%
	\put(0.5848382,0.44585602){\color[rgb]{0,0,0}\makebox(0,0)[lt]{\lineheight{1.25}\smash{\begin{tabular}[t]{l}$\omega$\end{tabular}}}}%
	\put(0.43830356,0.36912429){\color[rgb]{0,0,0}\makebox(0,0)[lt]{\lineheight{1.25}\smash{\begin{tabular}[t]{l}$\omega$\end{tabular}}}}%
	\put(0.21787915,0.70038081){\color[rgb]{0,0,0}\makebox(0,0)[lt]{\lineheight{1.25}\smash{\begin{tabular}[t]{l}North pole\end{tabular}}}}%
	\put(0,0){\includegraphics[width=\unitlength,page=4]{rotation_omega.pdf}}%
	\put(0.50578344,0.71531076){\color[rgb]{0,0,0}\makebox(0,0)[lt]{\lineheight{1.25}\smash{\begin{tabular}[t]{l}$\omega$\end{tabular}}}}%
	\end{picture}%
	\endgroup%
\end{itemize}

\section{Position of the sun relatively to the considered point}\label{s.positions_relatives}

In our basis, poin tat the earth's surface with latitude and longitude $0$, called null point (in Atalntic, $600$ kms at south of Accra, Ghana's capital) corresponds to $e_x$ vector, without obliquity, at day $\jr=0$ (i.e. at $21^{st}$ december) at noon ($\hr=0$ at UTC).
To obtain point with latitude $\lat$ and longitude $0$, we apply 
$\Rdeux_\lat$ rotation to the $e_x$ vector, then  $\Rtrois_\hr$ to obtain its position at hour $\hr$, which give us
$\Rtrois_\hr \times\Rdeux_\lat e_x$ 
Now, considering the inclination of earth's rotation axis, we apply 
 $\Rdeux_\obl$ rotation to the previous result : we obtain $\Rdeux_{\obl}\times \Rtrois_\hr \times \Rdeux_{\lat} e_x$.
 We are still cosidering the day of winter solstice, thus we will think to what happens at day $\jr$. 
 At that day, earth has the same orientation in heiocentric referential, but with a translation so that incoming sunbeam are not arising from $e_x$ direction, but  from $-\Rtrois_\jr e_x$ direction. As mentionned previously, the quantity of sunbeam intercepted by one horizontal square meter is proportional to the scalr product of the following vectors : $\overrightarrow{OM}=\Rdeux_{\obl}\times \Rtrois_\hr \times \Rdeux_{\lat} e_x$ and the unit vector of incoming sunbeam : $\Rtrois_\jr e_x$ (or $0$ if that quantity is non-positive). We know that applying the adoint of $\Rtrois_\jr$ matrix to $\Rdeux_{\obl}\times \Rtrois_\hr \times \Rdeux_{\lat} e_x$ leads to do the same computation as for 
 \[
 \bigmatrice_{\obl,\jr,\hr,\lat}=\Rtrois_{-\jr}\times \Rdeux_\obl \times \Rtrois_\hr\times\Rdeux_{\lat} \in SO_3(\R), \]
 the interesting quantity becomes $\bigmatrice_{\obl,\jr,\hr,\lat}e_x\cdot e_x$, that is, the coefficient on first line and first column of this product of $4$ rotations matrix. 
 Geometrically, the fact to apply $\Rtrois_\jr$ adjoint consists in considering that both erth and sun stay at their own place, but earth is subject to a rotation of angle $\jr$.
\[
\Rdeux_{\obl}\times \Rtrois_\hr \times \Rdeux_{\lat} e_x \cdot \Rtrois_\jr e_x
= \Rtrois_{-\jr}\times\Rdeux_{\obl}\times \Rtrois_\hr \times \Rdeux_{\lat} e_x \cdot  e_x.
\]
%

%

%
Since commutation with different axes do not commutate, expression of $\bigmatrice_{\obl,\jr,\hr,\lat}$ can not be simplified as it if were the case (in which case, we would simply obtain $\Rtrois_{-\jr+\hr}\times \Rdeux_{\obl+\lat}$)
%
Reality is (hopefully !) more complicated and first column of $\bigmatrice_{\obl,\jr,\hr,\lat}$, corresponding to $\bigmatrice_{\obl,\jr,\hr,\lat}e_x$ is

\[\begin{pmatrix}
\sin(\jr) \cos(\lat) \sin(\hr)+\cos(\jr) (\cos(\obl) \cos(\lat) \cos(\hr)-\sin(\obl) \sin(\lat)) \\
\cos(\jr) \cos(\lat) \sin(\hr)-\sin(\jr) (\cos(\obl) \cos(\lat) \cos(\hr)-\sin(\obl) \sin(\lat))\\
\sin(\obl) \cos(\lat) \cos(\hr)+\cos(\obl) \sin(\lat)
\end{pmatrix}	
\]

Second column, corresponding to $\bigmatrice_{\obl,\jr,\hr,\lat} e_y$, equals

\[\begin{pmatrix}
\sin(\jr) \cos(\hr)-\cos(\obl) \cos(\jr) \sin(\hr)  \\
\cos(\obl) \sin(\jr) \sin(\hr)+\cos(\jr) \cos(\hr) 	 \\
-\sin(\obl) \sin(\hr) 
\end{pmatrix}	
\]

And third column, corresponding to $\bigmatrice_{\obl,\jr,\hr,\lat} e_z$ equals

\[\begin{pmatrix}
\cos(\jr) (-\cos(\obl) \sin(\lat) \cos(\hr)-\sin(\obl) \cos(\lat))-\sin(\jr) \sin(\lat) \sin(\hr) \\
-\cos(\jr) \sin(\lat) \sin(\hr)-\sin(\jr) (-\cos(\obl) \sin(\lat) \cos(\hr)-\sin(\obl) \cos(\lat)) \\
\cos(\obl) \cos(\lat)-\sin(\obl) \sin(\lat) \cos(\hr)
\end{pmatrix}.
\]

\begin{rmk}[Why can we assume $\lambda = 0$ without loss of generality]
	Both obliquity, latitude, day and hour have an absolute meaning. On the contrary, the choice of a reference for longitude is an arbitrary convention : generally, Greenwhich's meridian (that is, point on earth's surface that belong to a plan containing North's ans South's poles, and going threw the city of Greenwhihc in the suburb of London, where we can find Royal Observatory). The refernce meridian for IERS (longitude $\lambda=0$) corresponds (up to few hundreds of meters) to that line. Any other point (except for the poles), could have been dercreted as being the reference, thus, in order to win in simplicity, the following only deals with longitude zero (without losing generality). 
 If we really want to know what happens for longitude  $\lambda\neq 0$ longitude, wehave to apply $\Rtrois_\lambda$ rotation at googd place, so that we obtain $\tilde\bigmatrice_{\obl,\jr,\hr,\lat,\lambda}:=\Rtrois_{-\jr}\times \Rdeux_\obl \times \Rtrois_\hr\Rtrois_\lambda\times\Rdeux_{\lat}$. Since $\Rtrois_\hr $ and $\Rtrois_\lambda$ share the same axis, they commutate, so that $\Rtrois_\hr\times \Rtrois_\lambda ) \Rtrois_\hr+\lambda$ : we find our previous result, that consideration only causes a time shift of a constant on a day. Indeed, point located on the same latitude observe the same phenomena with a delay relative to their respective longitude differences. for example, two such points with longitude $\lambda_1$ and $\lambda_2$ have a time shift of 
 $\frac{\lambda_1-\lambda_2}{2\pi}\times 24$ hours. 
\end{rmk}

\section{Computation of day length}\label{s.duree_jour}

Given $t, \lat$, observing variation over a day, we observe that there exists a  $\delta$-shift
such that $a\cos(\hr)+b\sin(\hr) = \sqrt{a^2+b^2} \cos(\hr+\delta)$, where $a=\cos(\obl)\cos(\lat)\cos(\jr)$ and $b= \cos(\lat)\sin(\jr)$. To find that $\delta$, one can evaluate for $\hr=0$ : we obtain $a\cos(0)+b\sin(0)=\sqrt{a^2+b^2}\cos(\delta)$. On can also obtain its sinus $\frac{\pi}{2}$ : from what we deduce $b=\cos(\frac{\pi}{2}+\delta)=-\sin(\delta)$ from where
$
\sin(\delta) = \frac{-b}{\sqrt{a^2+b^2}}.
$. 
Finally it is more convenient to use tangent (even if we lose information, due to the non injectivty, we will obtain a result valid only up to a constant times $\pi$ instead of $2\pi$).
For that purpose, we divide the two previous quantities :
$\tan(\delta) = \frac{-b}{a}$,that consideration avoid to keep the expression of square roots, which are not elegant. 
\begin{equation}\label{def.delta}
\delta \equiv \arctan\left(
\frac{-\sin(\jr)}{\cos(\obl)\cos(\jr)}\right) [\pi]
\equiv \arctan\left(
\frac{-\tan(\jr)}{\cos(\obl)}\right) [\pi]
\end{equation}

\begin{rmk}[If axis would not be inclinated]
	If $\obl =0$, then $\cos(\obl)=1$ so that argument in the arctangent becomes $-\tan(\jr)$, then time shift becomes $-\jr [\pi]$. This is reasonnable : after having done a quarter of loop around sun, the instant in the day in which sun attains it's summit (over the day) called noon, occurs one quarter of day later, that is, with a shift of $\frac{2\pi}{4}$ (we could replace in what precceds the $\frac 1 4 $ fraction by any other).
	Other remark : this time shift depends only on day and obliquité, so that, on a same meridian line, the time called noon happens simulateously.
\end{rmk}
We define the quantity $P_\obl(\lat,t,\hr)$ as the multiplying factor appearing in the quantity of power received by a section of one saure meter. We have
\[P_\obl(\lat,t,\hr)= \cos(\lat) \sqrt{\cos^2(\obl)\cos^2(\jr)+\sin^2(\jr)}\cos(\hr+\delta)-\sin(\obl)\sin(\lat)\cos(\jr).\]
Now, we denote by $p_\obl(\lat,t,\hr)$, the same quantity, shifted by  $\delta$ in its "hour" variable. $\hr$ : that transformation has the effect to center the day around noon : $\hr=0$.
\[
p_\obl(\lat,t,\hr) =   \cos(\lat) \sqrt{\cos^2(\obl)\cos^2(\jr)+\sin^2(\jr)}\cos(\hr)-\sin(\obl)\sin(\lat)\cos(\jr).
\]
\\
For  fixed $t,\lat$, the quantity inside parenthesis equals $0$ if and only if
 \[\cos(\lat) \sqrt{\cos^2(\obl)\cos^2(\jr)+\sin^2(\jr)}\cos(\hr)=\sin(\obl)\sin(\lat)\cos(\jr).\] 
That is possible (since $\cos(\hr)$ take all values in $[-1;1]$) if and only if
\[
\frac{\sin(\obl)\tan(\lat)\cos(\jr)}{\sqrt{\cos^2(\obl)\cos(\jr)+\sin^2(\jr)}}\in [-1;1].
\]
That is always the case for
 $\lat\in [0,\pi/2 -\obl]$ (that is between equator and arctic circle). For values of $\lat$ between
  $\pi/2 - \obl$ and $\pi/2$, that is possible only for day outside th period of olar day or polar night (described in \ref{s.nuit_jour_polaires}).
These pathologic values excepted, for general $(t,\lat)$ we have
\begin{eqnarray}\label{e.defp}
p_\obl(\lat,t,\hr)>0 &\Leftrightarrow & \cos(\hr) \ge \frac{ \sin(\obl)\sin(\lat)\cos(\jr)}{\cos(\lat) \sqrt{\cos^2(\obl)\cos^2(\jr)+\sin^2(\jr)}} \\
&\Leftrightarrow & \hr\in[-a_\obl(t,\lat),a_\obl(t,\lat)] 
\end{eqnarray}
from where
\begin{equation}\label{e.def.a_obl}
a_\obl(t,\lat)=\arccos\left(\frac{ \sin(\obl)\sin(\lat)\cos(\jr)}{\cos(\lat) \sqrt{\cos^2(\obl)\cos^2(\jr)+\sin^2(\jr)}}\right).
\end{equation}

Quantity $a_\obl(t,\lat)$ is the half of day length
 (expressed in radian, $2\pi = 24 h$), instant the sun is at summit corresponding to $\hr=0$, the considered point is lightened by the sun 
between $-a(t,\obl,\lat)$ and $+a(t,\obl,\lat)$ ( modulo $2\pi$).

\hspace{-2.8cm}\includegraphics[scale=0.5]{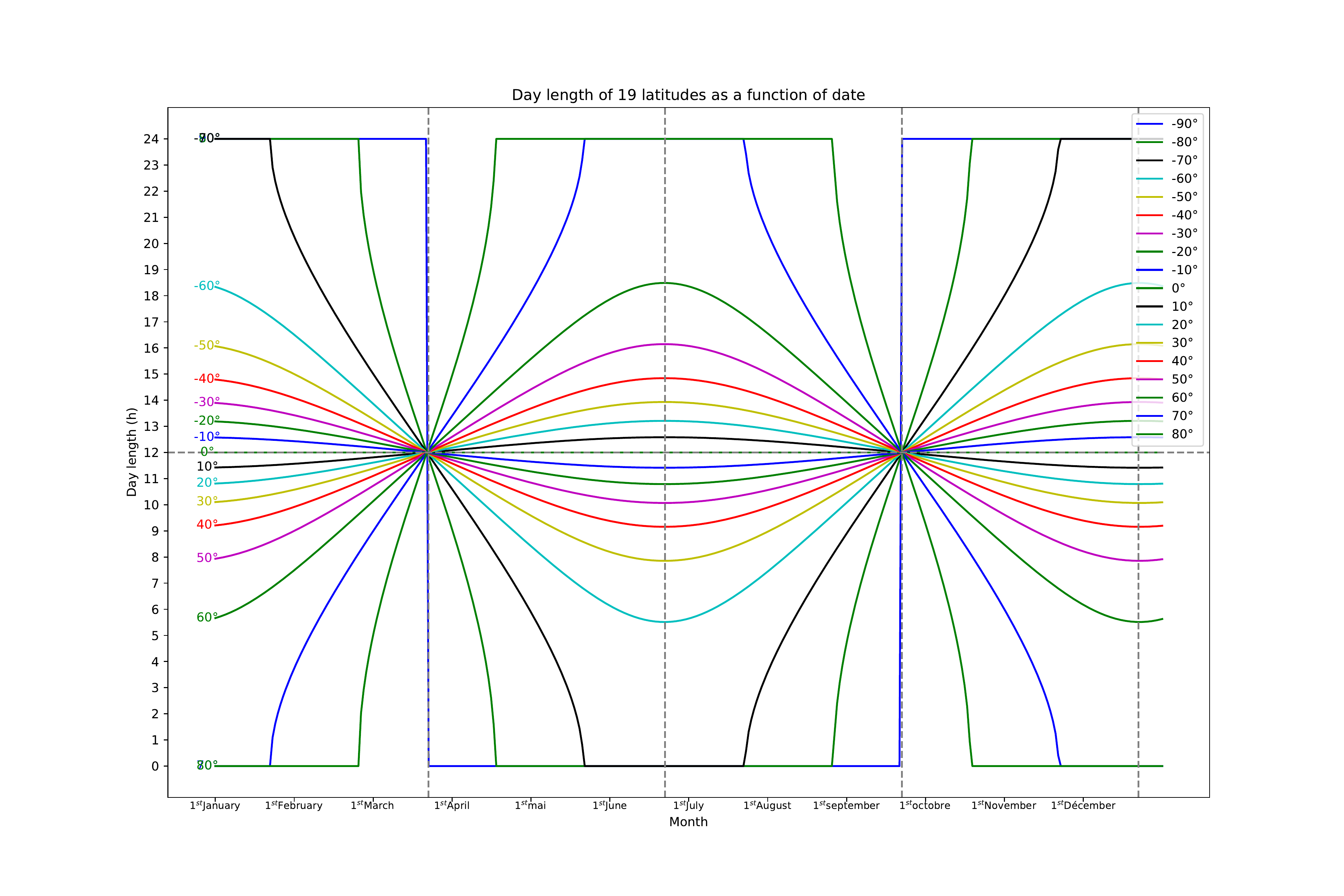}
\begin{rmk}[Remark on the day length]
	\begin{itemize}
		\item \emph{At equator}. 
		For a point belonging to equator, we have
	$\lat=0$ thus $\sin(\lat)=0$ so that quantity to whicharccosinus is applied equals $0$. Hence
 $a_\obl(t,0)$ is constannt to  $\frac{\pi}{2}$ so that day length equals $12h$, at any day in the year.
		
		\item \emph{If earth would not be leaning} 
		If $\obl=0$, then
	 $\sin(\obl)=0$ 
	 and for same reasons as previous remark, day length would be on any point at any day equals to $12h$.
		
		\item \emph{At north pole.} We have $\cos(\lat)=0$ 
		
		so that expression to which arccosinus is applied would be
		 $+\infty$ or $-\infty$. 
		 Actually, for these two very specifics cases, position is constant over a day, which last either $0h$ (between autumn and spring equinoxes for north) or $24h$ (the opposite) (and the oposite for south's pole).

		\item \emph{At spring equinox (resp. autumn)}, which happens around $21^{st}$ of septemberwe have $t_{eq.spring}=\frac{\pi}{2}$ (resp. $t_{eq.autumn}=\frac{3\pi}{2}$) thus $\cos(t_{eq.printemps})=\cos(t_{eq.automne})=0$. Then, at any latitude $\lat$, day length equals
		\[
		a_\obl(t_{eq.spring},\lat)=a_\obl(t_{eq.autumn},\lat)=\arccos\left(0\right)=\pi/2.
		\]
		For this specific day, on any point on earth's surface, we can observe a day of $12h$ and a night of $12h$.

		\item \emph{Antipodic points.}
		Two points are antipodic if they have opposite latitude and longitude $\lat_1=-\lat_2$ and $\lambda_1=-\lambda_2$. Since the delimitation between day and night zones is realised by the plan cutting erth in two, orthogonal to sun direction, and containing two poles, we claim that for almost all time, exactly one of the two antipodic points is lightened by the sun, while the other is plunged into the night. 
		Thus, if we add all over a year the day length of these two points, we obtain $1 $ year, idem for night. Since two point with opposite latitude share the same behaviour but with $6$ months time shift : $a_\obl(\jr,\lat) = a_\obl(\jr+\pi,-\lat)$ we know that the contributin of point $1$ and $2$ are the same, so that, any point on earth is lightened by sun half-time, and in the shadow the other half time. Moreover, we have 
		$a_\obl(\jr,\lat)=2\pi-a_\obl(\jr,-\lat)=2\pi-a_\obl(\jr,\lat)$.
	\end{itemize}
\end{rmk}

\begin{rmk}[Model's limits]
	
	\begin{itemize}
		\item Earth's orbit is not circular, but elliptic. Its eccentricity is sufficiently weak to make its assumption reasonnable. 
		\item 
		Obliquity is not constant, but is subjet to nutation phenomenon. 
		\item Earth is not exactly spherical.
		\item Sun is not realy a point at infiny, sunbeams are not realy parralel due to that fact.
		\item 
		Even when sun is completely under horizon, luminosity can still persists, by phenomen of diffusoin. Indeed, the presence of atmosphere diffuses sunbeams, mostly those with lower frequency (red). When sun is under $0$ and $6$ degrees, civil twilight occurs, between $6^\circ$ and $12^\circ$, nautical twilight occurs, and between $12^\circ$ and $18^\circ$, astronomical twilight occurs. 
		To answer previous questions with for example civil night, unstead of solving equation with right member equals to $0$, we replace this last by $\sin\left( \frac{6}{360}2\pi \right)$ (resp. $\sin\left( \frac{12}{360}2\pi \right)$ or $\sin\left( \frac{18}{360}2\pi \right)$ for nautical and astronomical twilight).
	\end{itemize}
\end{rmk}

\section{Polar day and polar night}\label{s.nuit_jour_polaires}
It is night, acording to \eqref{e.defp} if and only if $p_\cdot(\cdot,\cdot,\hr)<0$, that is, for given $\lat,t$ if and only if 
\[	 \cos(\hr) \le \frac{ \sin(\obl)\sin(\lat)\cos(\jr)}{\cos(\lat) \sqrt{\cos^2(\obl)\cos^2(\jr)+\sin^2(\jr)}} 
\] 
Polar night occurs at latitude
 $\lat$ at day $\jr$ if and only if maximum of 
 $p_\obl(\lat,\jr,\cdot)$ over a day, which is reached for $\hr=0$ (because $\cos(0)=1$), is lower than the right member of previous inequality
i.e. if and only if 
$\sin(\lat)\cos(\jr)>0$  and

\begin{eqnarray*}
	&\cos^2(\lat) \left(\cos^2(\obl)\cos^2(\jr)+\sin^2(\jr)\right)  \leq \sin^2(\obl)\sin^2(\lat)\cos^2(\jr)   \\
	\Longleftrightarrow&
	\left(\cos^2(\obl)\cos^2(\lat) -\sin^2(\obl)\sin^2(\lat)\right) \cos^2(\jr) \leq  -\cos^2(\lat)\sin^2(\jr)
	\\
	\Longleftrightarrow&
	\left(\cos^2(\obl)\cos^2(\lat) -\sin^2(\obl)\sin^2(\lat)\right) \leq  -\cos^2(\lat)\tan^2(\jr)\\
	\Longleftrightarrow&
	\frac{\sin^2(\obl)\sin^2(\lat)-\cos^2(\obl)\cos^2(\lat)}{\cos^2(\lat)}
	\geq  \tan^2(\jr)
\end{eqnarray*}
Since the square of tangent is non-negative, that is possible only if
\[\frac{\sin^2(\obl)\sin^2(\lat)-\cos^2(\obl)\cos^2(\lat)}{\cos^2(\lat)}\geq 0,\] i.e.  \ssi
$\cos^2(\obl)\cos^2(\lat) -\sin^2(\obl)\sin^2(\lat)\leq 0 $ ou encore $\cos^2(\obl)\cos^2(\lat) \leq \sin^2(\obl)\sin^2(\lat) $. 
What precedes is equivalent to $\cotan^2(\lat)=\tan^2(\frac \pi 2 - \lat)\leq \tan^2(\obl)$. Considering latitudes of north's hemisphere, $\lat\in[0,\frac\pi 2]$, we have $\tan(\lat)>0$ which is equivalent to
$\lat \in[\frac \pi 2 -\obl,\frac \pi 2]$. 
The polar night phenomenon could happens, in north hemisphere, only between arctic (from greek $\alpha\rho\kappa\tau o\varsigma$  (árktos) meaning bear) circle and north's pole. That corresponds to latitude between 
 $\frac{\pi}{2}-\obl$ and $\frac \pi 2$. 
 
 When one of these latitudes $ \phi $ is plunged into polar night, the latitude $ - \phi \in [- \frac \pi 2, - \frac \pi 2 + \obl] $ is bathed in polar day, and Conversely. When the polar night reigns in one of these latitude, we can also guarantee that 6 months later ($ \jr_2 = \pi + \jr_1 $) the polar day will prevail and vice versa. Finally, in one of these given latitudes, the start and end dates of the polar night (respectively of the polar day) are symmetrical with respect to the winter solstice (respectively: summer). The duration of this polar day and night are identical, increasing from $ 0 $ days in the polar circle (the critical case) to $ 6 $ months in the pole (the extreme case).
 
Now consider a latitude $\lat $ lying north of the Arctic Circle (by symmetry, the calculations for $ - \lat $ lying south of the Antarctic Circle would be similar). As we have just seen, the polar night takes place if and only if

$\frac{\cos^2(\obl)\cos^2(\lat) -\sin^2(\obl)\sin^2(\lat)}{\cos^2(\lat)}
\geq  \tan^2(\jr)$ and the sign condition $\sin(\lat)\cos(\jr)>0$, 
and in north's hemisphere
$\sin(\lat)>0$ and $\cos(\jr)>0$ means that we are more close to winter solstice than to summer solstice. Assuming these conditions, polar night happens only for days  $\jr$ satisfying
\[
\jr \le \arctan\left(  \sqrt{\cos^2(\obl)\cos^2(\lat) -\sin^2(\obl)\tan^2(\lat)} \right).
\]
Thus, noting  $N(\obl,\lat)=\arctan\left(  \sqrt{\cos^2(\obl)\cos^2(\lat) -\sin^2(\obl)\tan^2(\lat)} \right)$, we have, for any $\lat \in[ \frac \pi 2 - \obl, \frac \pi 2]$,
\begin{itemize}
	\item Polar night happens for $\jr \in [-N(\obl,\lat),N(\obl,\lat)]$.
	\item Polar day happens for $\jr \in [\pi-N(\obl,\lat),\pi+N(\obl,\lat)]$.
\end{itemize}
That is, latitude has to belong to the intervall 
 $[\frac{\pi}{2}-\obl,\frac{\pi}{2}]$,or its opposite, which corresponds to zone between antarctic circle and south's pole.
For these latitudes; beginning of polar night happens for
\[
t_{beginning~polar~night}(\obl,\lat)=\arctan\left( 
\frac{\cos^2(\lat)}{\sin^2(\lat)-\cos^2(\obl)}
\right).
\]
One can also fix a day (between spring and autumn equinoxes) and ask 
On peut aussi se fixer une date (entre les équinoxes d'automne et de printemps) et se demander north of which latitude it is constantly dark (taking an opposite latitude, this amounts to wondering: south of which latitude, it is constantly day).
Given a day $\jr$, it is polar night at latitude $\lat$ if and only if :
\[
\cotan(\jr)^2 \leq 
\frac{\sin^2(\lat)-\cos^2(\obl)}{\cos^2(\lat)},
\]
which comes down to
\[
\cos^2(\lat)\frac{\cos^2(\jr)}{\sin^2(\jr)}\le  \sin^2(\lat) - \cos^2(\obl) \Leftrightarrow
\cos^2(\obl)+\frac{\cos^2(\jr)}{ \sin^2(\jr)}\le \sin^2(\lat)\left( 1+\frac{\cos^2(\jr)}{\sin^2(\jr)} \right)
\]
that is
\[
\sin^2(\lat)
\ge 
\sin^2(\jr)\cos^2(\obl)+\cos^2(\jr)
\]
That corresponds to north of latitude
\[
\lat_{polar~night~limit}(\obl,\jr) = \arcsin\left( \sqrt{\sin^2(\jr)\cos^2(\obl)+\cos^2(\jr)} \right).
\]
Since $\sin^2(\jr)\cos^2(\obl)+\cos^2(\jr)=\sin^2(\jr)\cos^2(\obl)+1-\sin^2(\jr)
=\sin^2(\jr)(\cos^2(\obl)-1)+1$ we have
\[
\lat_{polar~night~limit}(\obl,\jr) = \arcsin\left( \sqrt{1-\sin^2(\obl)\sin^2(\jr)} \right).
\]

\hspace{-3cm}\includegraphics[scale=0.5]{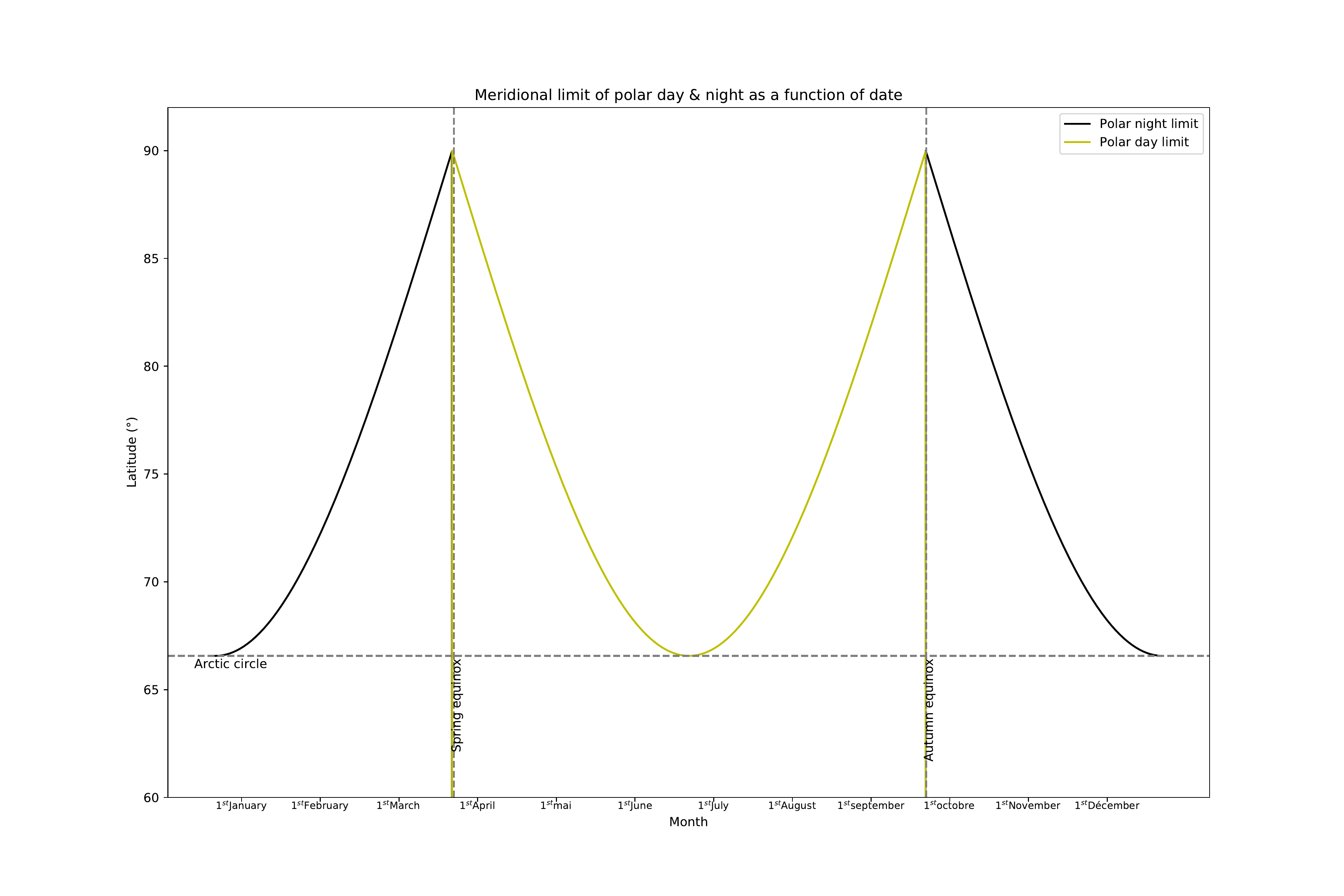}
\begin{rmk}[Petite anecdote personnelle]
	Il m'est arrivé de voyager à bicyclette au nord du cercle polaire à une période entre le solstice d'été et l'équinoxe d'automne (début août, donc approximativement à mi-chemin entre les deux). Vivre un jour ou une nuit polaire faisant partie de mes principales volontés, la question que je me posais, par exmple lorsque nous nous trouvions au $68^{eme}$ parallèle Nord (donc à peine plus que $66.7$ le cas critique)là où le ferry nous a déposés à Svolsvaer,à une date donnée, à quelle vitesse il fallait remonter (en supposant que la route remonte plein Nord ce qui n'est pas le cas sur le très sinueux littoral norvégien !) pour dépasser cette ligne. N'ayant pas encore fait tous ces calculs, en première approximation (en bon physicien quoi), on peut estimer que la latitude limite de jour polaire se trouve à la latitude $66.7^\circ$ au 21 juin et remonte à vitesse constante (c'est là qu'est bien sûr l'approximation) pour se trouver à $90^\circ$ nord le $21$ juin, soit $23^\circ26'=23+\frac{26}{60} ^\circ$ en trois mois, i.e. $91,3$ jours.
	Ceci correspond donc à une distance de $\frac{23+\frac{26}{60}}{360}\times 40 000$ en $91$ jours soit $28,6 km/jour$. Autrement dit, largement rattrapable en faisant un petit effort ! La vitesse calculée correspond en fait à la vitesse moyenne de la délimitation  ``nuit polaire'', pour connaître sa vitesse instantanée, il faut considérer sa dérivée en la variable $\jr$ qui est
	\[
	\frac{\partial}{\partial_t}\lat_{m\acute{e}ridionnale~de~nuit~polaire}(\obl,\jr) = \frac{\partial}{\partial_t} \arcsin\left( \sqrt{\sin^2(\jr)\cos^2(\obl)+\cos^2(\jr)} \right)
	\]
	On calcule cette dérivée partielle et l'on trouve
	\[ 
	\frac{-\sin^2(\obl)\cos(\jr)\sin(\jr)}{\sqrt{\sin^2(\jr)\cos^2(\obl)+\cos^2(\jr)}\sqrt{1-\sin^2(\jr)\cos^2(\obl)+\cos^2(\jr) }}.
	\]
	Le théorème de Rolle nous apprend qu'il y a au moins un instant auquel la vitesse sera bien de $28.6 km/jour$, et en fait l'approximation fournit un ordre de grandeur convaincant. Pour conclure sur la petite anecdote : nous n'avons pas rattrapé cette ligne, mais grâce à la diffusion de la lumière, il y avait de la luminosité à toute heure du jour (crépuscule polaire).

\end{rmk}

\section{Solar energy received over a day}\label{s.energiejour}
If we want to know the amount of energy received by a horizontal surface of one square meter during a day at the time of the year $ \jr $ at latitude $ \lat $, it suffices to integrate between $ 0 $ and $ 2 \pi $ the quantity $ p_\obl (\lat, t, \hr) $, but in order to get rid of the positive part, we can simply integrate between $ -a (t, \obl , \lat) $ and $ a (t, \obl, \lat) $. By noting this quantity $ J (\lat, t) $, we therefore have:
%
\begin{eqnarray*}
	J_\obl(\lat,t) &=& \int_{-a(t,\obl,\lat)}^{a(t,\obl,\lat)}  \cos(\lat)  \sqrt{\cos^2(\obl)\cos^2(\jr)+\sin^2(\jr)}\cos(\hr)\\& &-\sin(\obl)\sin(\lat)\cos(\jr) \,d\hr \\
	& = & 2 \cos(\lat) \sqrt{\cos^2(\obl)\cos^2(\jr)+\sin^2(\jr)}  \sin(a(t,\obl,\lat))\\& & - 2a(t,\obl,\lat)\sin(\obl)\sin(\lat)\cos(\jr).
\end{eqnarray*}
Considering that $a_\obl(t,\lat) = \arccos\left(\frac{ \sin(\obl)\sin(\lat)\cos(\jr)}{\cos(\lat) \sqrt{\cos^2(\obl)\cos^2(\jr)+\sin^2(\jr)}}\right)$ and that $\sin(\arccos(u)) = \sqrt{1-u^2}$, 
\begin{equation*}
\begin{split}
J_\obl(\lat,t) &= 2\cos(\lat) \sqrt{\cos^2(\obl)\cos^2(\jr)+\sin^2(\jr)} 
\sqrt{1 - \frac{ {\sin^2(\jr)\sin^2(\obl)\sin^2(\lat)}}{\cos^2(\lat) \left(\cos^2(\obl)\cos^2(\jr)+\sin^2(\jr)\right)}}   \\
&-2 \sin(\obl)\sin(\lat)\cos(\jr) \arccos\left( \frac{ \sin(\obl)\sin(\lat)\cos(\jr)}{\cos(\lat) \sqrt{\cos^2(\obl)\cos^2(\jr)+\sin^2(\jr)}} \right)\\
&= 2
\sqrt{\cos^2(\lat)\left(\cos^2(\obl)\cos^2(\jr)+\sin^2(\jr)\right) -  {\sin^2(\jr)\sin^2(\obl)\sin^2(\lat)}}  \\
&-2 \sin(\obl)\sin(\lat)\cos(\jr) \arccos\left( \frac{ \sin(\obl)\sin(\lat)\cos(\jr)}{\cos(\lat) \sqrt{\cos^2(\obl)\cos^2(\jr)+\sin^2(\jr)}} \right)
\end{split}
\end{equation*}
in fact this quantity is valid only if $ \arccos (...) $ is well defined (ie there is a problem during the night and the polar day), but by extending the definition of the arc-cosine in setting $ \arccos (x) = 0 $ if $ x> 1 $ (ie the duration of the day is worth $ 0 $ nevertheless at most of the day, the quantity $ P_\obl (\lat, \hr, t) $ is negative) and $ \arccos (x) = \pi $ if $ x <-1 $ (if the minimum of $ P_\obl (\lat, \hr, t) $ on the day is positive, ie for the day polar, we take $ a (\lat, t) = \pi $), the calculation remains valid.
 \hspace{-4cm}\includegraphics[scale=0.4]{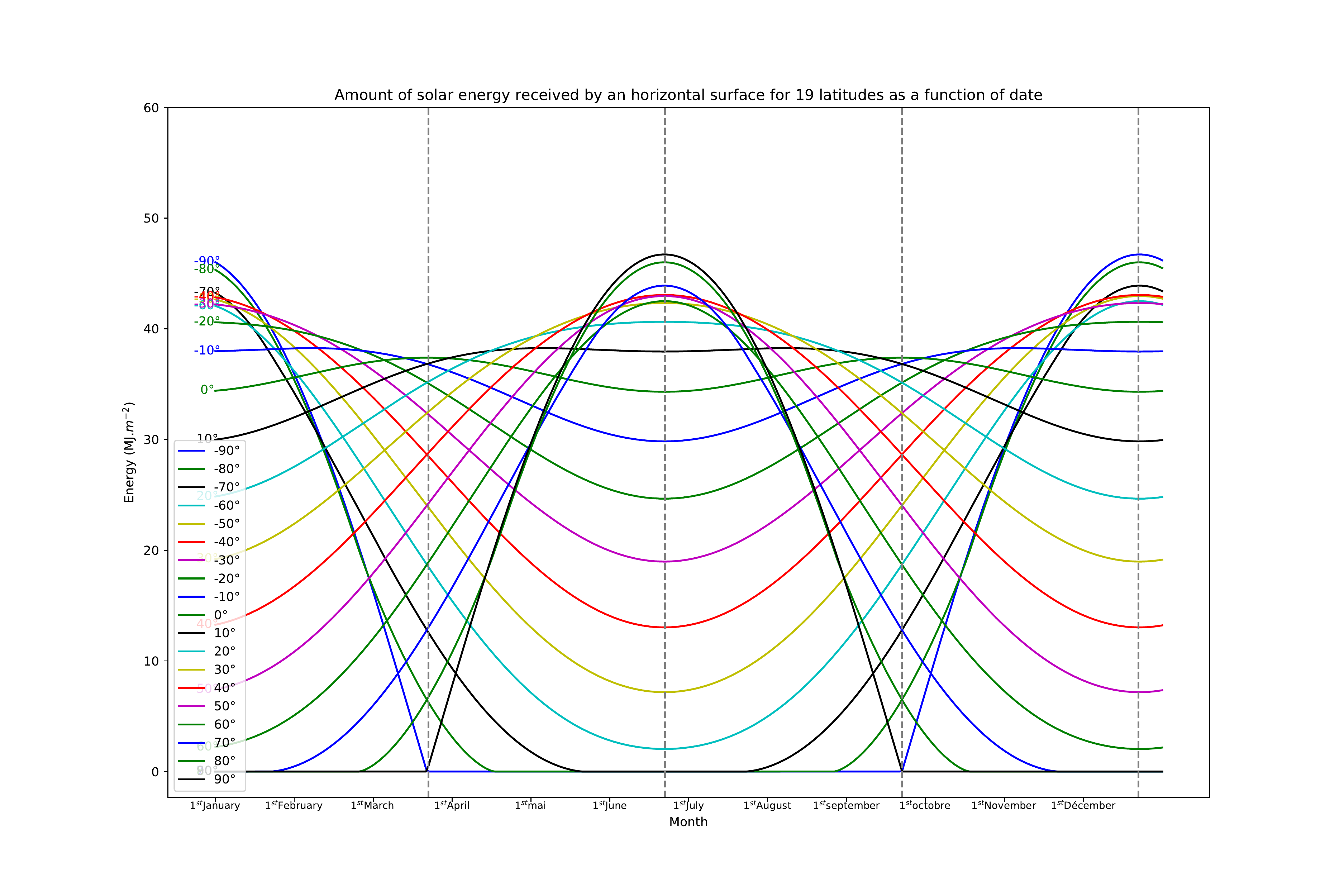}
\begin{rmk}[If earth would not be leaning]
If $ \obl = 0 $, we would have $ p_0 (\lat, t, \hr) = \cos (\lat) \cos (\hr) $. Which is an expression that does not depend on the day considered (no seasonal phenomenon without obliquity). By integrating this between $ - \frac \pi 2 $ and $ \frac \pi 2 $ (because the duration of the day is then constant being worth $ 12h $) we get $ 2 \cos (\lat) $: the surface energy received the greater the distance from the equator, whatever the day of the year. In addition, without obliquity, the polar zones do not receive any solar energy.
\end{rmk}

\begin{rmk}[At equator]
For latitude $\lat=0$, we have 
	\[p_\obl(0,t,\hr)=\sqrt{(\cos^2(\obl))\sin^2(\jr)+\cos^2(\jr)} \cos(\hr).\]
	Integrating it over a day, we obtain
	$J_\obl(0,\jr) = 2\sqrt{(\cos^2(\obl))\sin^2(\jr)+\cos^2(\jr)}$.
\end{rmk}

\begin{rmk}[At  poles]
	For latitude $\lat=\frac \pi 2$ (for $-\frac \pi 2$, consequences are anolous), we have $p_\obl(\frac \pi 2,t,\hr)= -\sin(\obl)\cos(\jr)$. Integrating over a day, we have
	$-2\pi\sin(\obl)\mathbf{1}_{\sin(\obl)<0}$ : 
	indeed, during a day, the position of the sun is constant. We realize again that the energy received from the sun by the polar zones is only due to the obliquity, because it is proportional to its sine, which as a first approximation is worth $\obl+o(\obl)$.
\end{rmk}

\begin{rmk}[Surprising maximum value]
	Given an obliquity $\obl$,one can look for the couple (latitude,day) which maximize $J_\obl(\lat,\jr)$ function, by researching its critical points.
	that is, values of $(\obl,\lat)$ satisfying
	$\frac{\partial}{\partial \lat} J_\obl(\lat,\jr) = \frac{\partial}{\partial \jr} J_\obl(\lat,\jr) = 0$. 
It would then suffice to compare the finite number of values obtained, because the maximum is necessarily reached at one of these critical points. The calculation of these partial derivatives is left to the reader : it is possible to use the integral formulation of 
$J_\obl$ 
and to apply the derivation theorem under the unequal symbol, taking care to note that the bounds are not fixed.
Without calculation, we can be convinced of one thing: north of the tropic of Cancer, for a given latitude $ \lat > \obl $, the day maximizing the value of $J_\obl (\lat, \cdot)$ corresponds at the summer solstice : $ \jr = \pi $. We have
	\[
	J_\obl(\lat,\pi) 
	=2\cos(\obl)\cos(\lat)-2 \sin(\obl)\sin(\lat) \arccos\left( - \tan(\obl)\tan(\lat)\right).	\]
	Derivating with respect to latitude, we obtain
	\begin{eqnarray*}
		\frac{\partial}{\partial\lat}
		J_\obl(\lat,\pi)  
		&=&-2\cos(\obl)\sin(\lat)+2 \sin(\obl)\times\\
		& &\left(\cos(\lat) \arccos\left( - \tan(\obl)\tan(\lat)\right)
		-\sin(\lat) \frac{\tan(\obl)}{\cos^2(\lat)} \frac{-1}{\sqrt{1-\tan^2(\obl)\tan^2(\lat)}}
		\right).
	\end{eqnarray*}
We could then be interested in solving the equation $ \frac{\partial}{\partial \lat}
J_\obl (\lat, \pi) = 0 $, but we can also see, graphically, that the maximum is achieved for the latitude $ \frac{\pi}{2} $ (with an obliquity of $ 23^\circ $), that a local minimum is around $ 7 ^\circ $ and a local (non-global) maximum around $ 40^\circ $.
\end{rmk}

\hspace{-2cm}
		\begin{tabular}{cc}
			\includegraphics[scale=0.2]{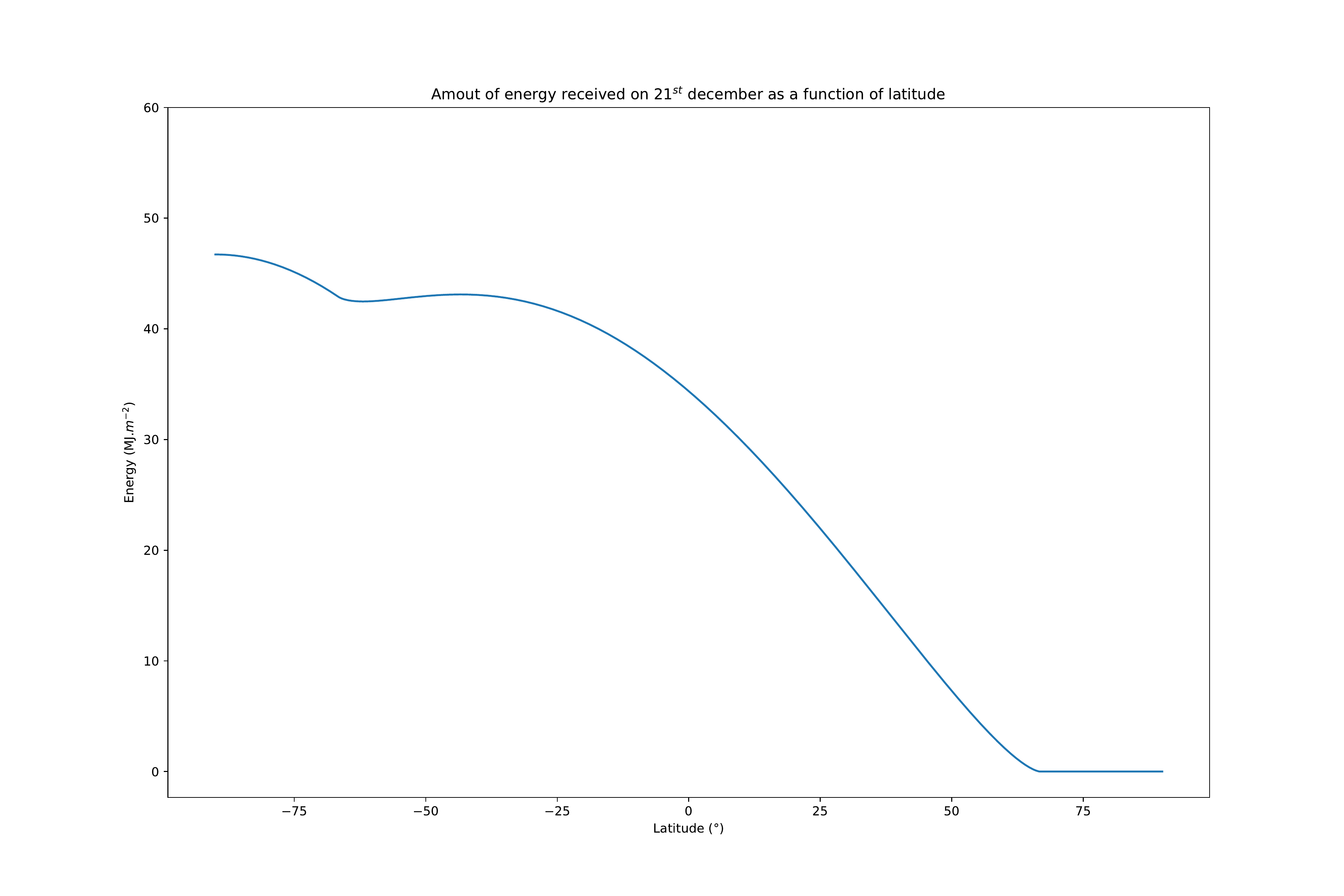} &
			\includegraphics[scale=0.2]{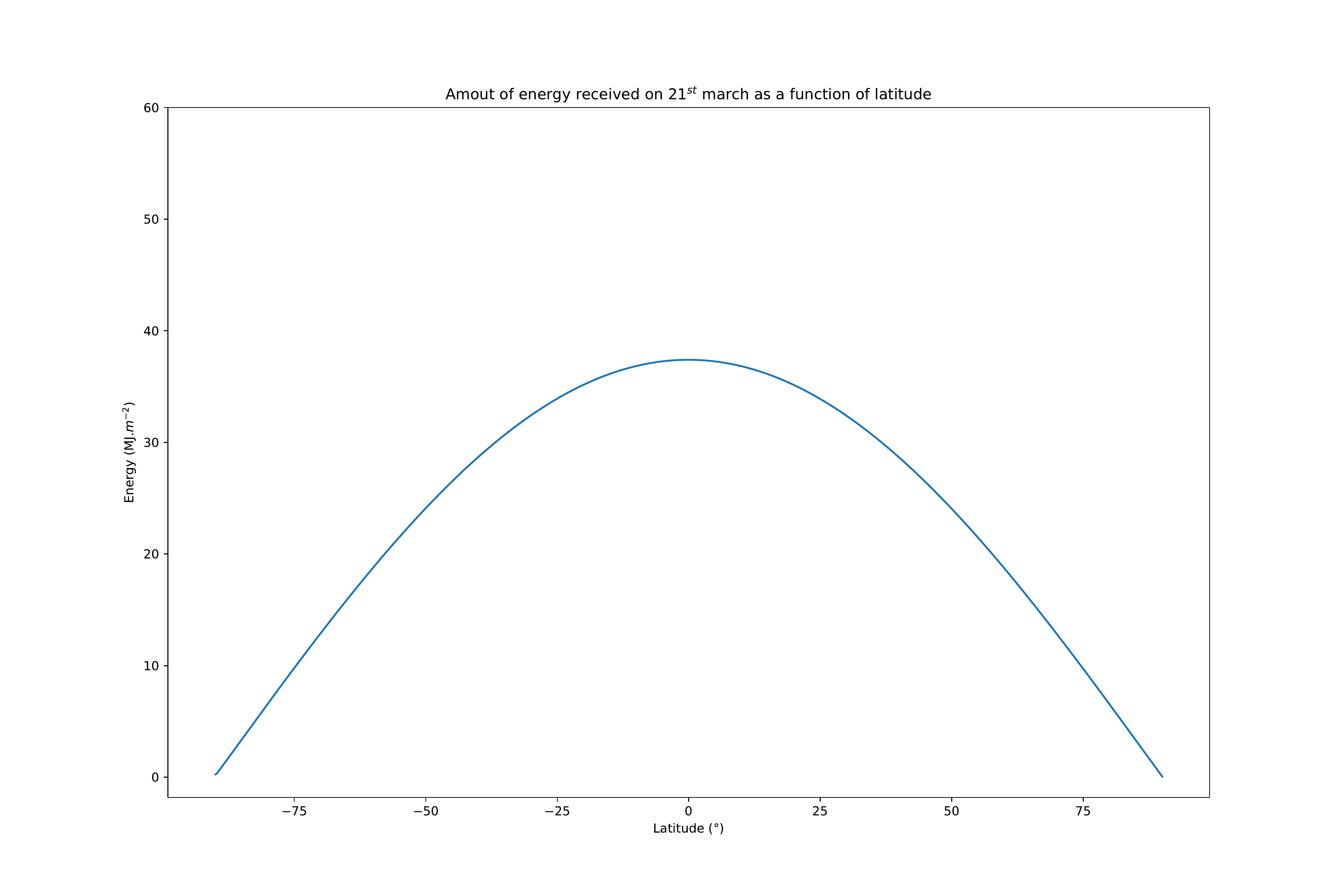}\\
		\end{tabular}

\begin{rmk}[Energy received by the entire surface of the earth in one day]
If at $\obl $ and $ \jr $ fixed, we realize the integral on the earth's surface and over a period of $ 24 $ 
consecutives hours of the power received from the sun, this corresponds to the integral of $ J_\obl ( \lat, \jr) $
 weighted by the area represented by the latitude $ \lat $: in fact, the area between latitudes
$ 0 $ and $\delta \lat $ represents a wide band (along the equator)
 of length $ 40,000 km $ and the width $ \delta \lat $, while the area
between latitudes $ \frac \pi 2- \delta \lat $ and $ \frac \pi 2 $
 represents only a circle of radius $ \delta \lat $ around the pole.
	\[
	\pi \int_{-\frac \pi 2}^{\frac \pi 2} \cos(\lat) J_\obl(\lat,\jr) \,d\lat.
	\]
	This quantity is constant, depends neither on the obliquity $ \obl $ nor on the date $ \jr $ because we can affirm, that by cumulating all the zones exposed to the sun in a given instant, we find the hemisphere exposed to the sun, which offers a constant `` useful '' surface (intercepting the rays of the sun) equal to $ \pi $ (or $ \pi R_{Earth} ^ 2 $ if we want units). We therefore obtain a constant value.
\end{rmk}

\begin{rmk}[Pour obtenir la véritable valeur de l'énergie perçue]\label{rem.fluxsolaire}
	So far, all calculated quantities are dimensionless. These are actually multiplicative factors.
	To make the link between $ p (\obl, \lat, \jr \hr) $ and the power received by a horizontal section of one square meter at latitude $ \lat $, at day $ \jr $ at hour $ \hr $ (with an obliquity $ \obl $), so just multiply by the solar constancy, expressed in $ 1361 Wm ^ {- 2} $. To obtain the real energy received on a day, it is necessary to integrate this quantity over a time interval of $ 24h $: this corresponds to multiply by $ \frac {24 \times 3600}{2 \pi} s $, so in fine , multiply
	 $J_\obl(\lat,\jr,\hr)$ by $1361 W.m^{-2}\times \frac{24\times 3600}{2 \pi}s \sim 18,7 MJ.m^{-2} $.
\end{rmk}

\section{Amount of energy received in a year}\label{s.energie_annee}

Using the expression obtained in \eqref{e.Pphi} 
and trying to integrate it over one year  (that is $[0,2\pi]\times[0,2\pi]$) we can also write
 $P_\obl(...,t)$ under shape $a\cos(\jr)+b\sin(\jr)$.
\begin{equation*}\label{e.Pphi}
P_\obl(\lat,t,\hr)= \left[\left(\cos(\obl)\cos(\lat)\cos(\hr)-\sin(\obl)\sin(\lat)\right)\sin(\jr)+\cos(\lat)\sin(\hr)\cos(\jr)\right]^+
\end{equation*}
As previously seen, there exists a shift
$\delta_\jr$whose expression plays no role but convenient to simplify the expression of $(A\cos(\jr)+B\sin(\jr))^+$. Indeed :
$(A\cos(\jr)+B\sin(\jr))^+=\sqrt{A^2+B^2}\cos(\jr+\delta_t)^+$. 
As what interests us below corresponds to an integral of this quantity on the torus, the change of variable $T= \jr+\delta_t$ will be convenient.
Here $A=\cos(\obl) \cos(\lat) \cos(\hr)-\sin(\obl) \sin(\lat)$ and $B=\cos(\lat) \sin(\hr)$

We note $A_\obl(\lat)$ the multiplicative factor (to be multiplied by the solar flux) giving the solar energy received by a square meter at latitude $ \lat $ in a model where the axis of the earth has an obliquity $ \obl $:

\begin{eqnarray*}
	A_\obl(\lat) 
	&=&\int_{0}^{2\pi} \int_{0}^{2\pi} \left(A_{\obl,_lat,\hr}\cos(\jr)+B_{\obl,\lat,\hr}\sin(\jr)\right)^+ \,d\hr \,d\jr \\
	&=&\int_{0}^{2\pi} \int_{0}^{2\pi} \sqrt{A_{\obl,\lat,\hr}^2+B_{\obl,\lat,\hr}^2} \cos(\jr+\delta_t)^+ ,d\hr \,d\jr\\
	&=&\int_{0}^{2\pi}\cos(\jr+\delta_t)^+ \, d\jr \int_{0}^{2\pi}\sqrt{A_{\obl,\lat,\hr}^2+B_{\obl,\lat,\hr}^2} \,d\hr \\
	&=&\int_{-\frac \pi 2}^{\frac \pi 2} \cos(T)\,dT \int_{0}^{2\pi}\sqrt{A_{\obl,\lat,\hr}^2+B_{\obl,\lat,\hr}^2} \,d\hr\\
	&=& 2 \int_{0}^{2\pi}\sqrt{A_{\obl,\lat,\hr}^2+B_{\obl,\lat,\hr}^2} \,d\hr
\end{eqnarray*}
Thus, replacing the respective expressions $A_{\obl,\lat,\hr}$ and $B_{\obl,\lat,\hr}$ by their actual values
\begin{equation*}
A_\obl(\lat) = 2\cos(\lat)
\int_{0}^{2\pi}\sqrt{
	\left(
	\cos(\obl)  \cos(\hr)-\sin(\obl) \tan(\lat)
	\right)^2+\sin^2(\hr)
}
\,d\hr
\end{equation*}
This integral is performed on $ [0.2 \pi] $, but it could equally be done on any interval of length $ 2 \pi $ (in fact, it is integrated on the torus $ \mathbb {T} = \R / (2 \pi \Z) $). To exploit the parity of the cosine function, we can for example perform it on the interval $ [- \pi, \pi] $, then use the parity to affirm that the result is twice the integral on $ [0 , \pi] $. So we find, replacing $ \sin ^ 2 (\hr) $ by
 $1-\cos^2(\hr)$ :
\[
A_\obl(\lat) = 4\cos(\lat)
\int_{0}^{\pi}\sqrt{
	\left(
	\cos(\obl)  \cos(\hr)-\sin(\obl) \tan(\lat)
	\right)^2+1-\cos^2(\hr)
}
\,d\hr
\]
The cosine function performs a bijection from the interval $ [0, \pi] $ to the interval $ [- 1,1] $, by setting $ u = \cos (\hr) $, we verify that $ du = - \sin (\hr) d \hr = - \sqrt {1-u ^ 2} d \hr $ and so we have

%
\[
A_\obl(\lat) = 4\cos(\lat)
\int_{0}^{1}\sqrt{
	\frac{\left(
		\cos(\obl)  u-\sin(\obl) \tan(\lat)
		\right)^2+1-u^2}{1-u^2}
}
\,du.
\]
\begin{rmk}[To obtain actual value of received energy]
	In the same way as for the remark \ref{rem.fluxsolaire}, to establish the link between the real energy, expressed in $ Jm ^ {- 2} $ and the coefficient obtained, it is necessary to make correspond the set of integration $ [0.2 \pi] ^ 2 $ with $ 24h \times 365.25 days $ and multiply by the solar constant. This therefore corresponds to multiplying by $\frac{365.25\times24\times3600\times 1361}{4\pi^2}J.m^{-2}$, i.e nearly $1,09 GJ.m^{-2}$.
\end{rmk}
\hspace{-2cm}\includegraphics[scale=0.5]{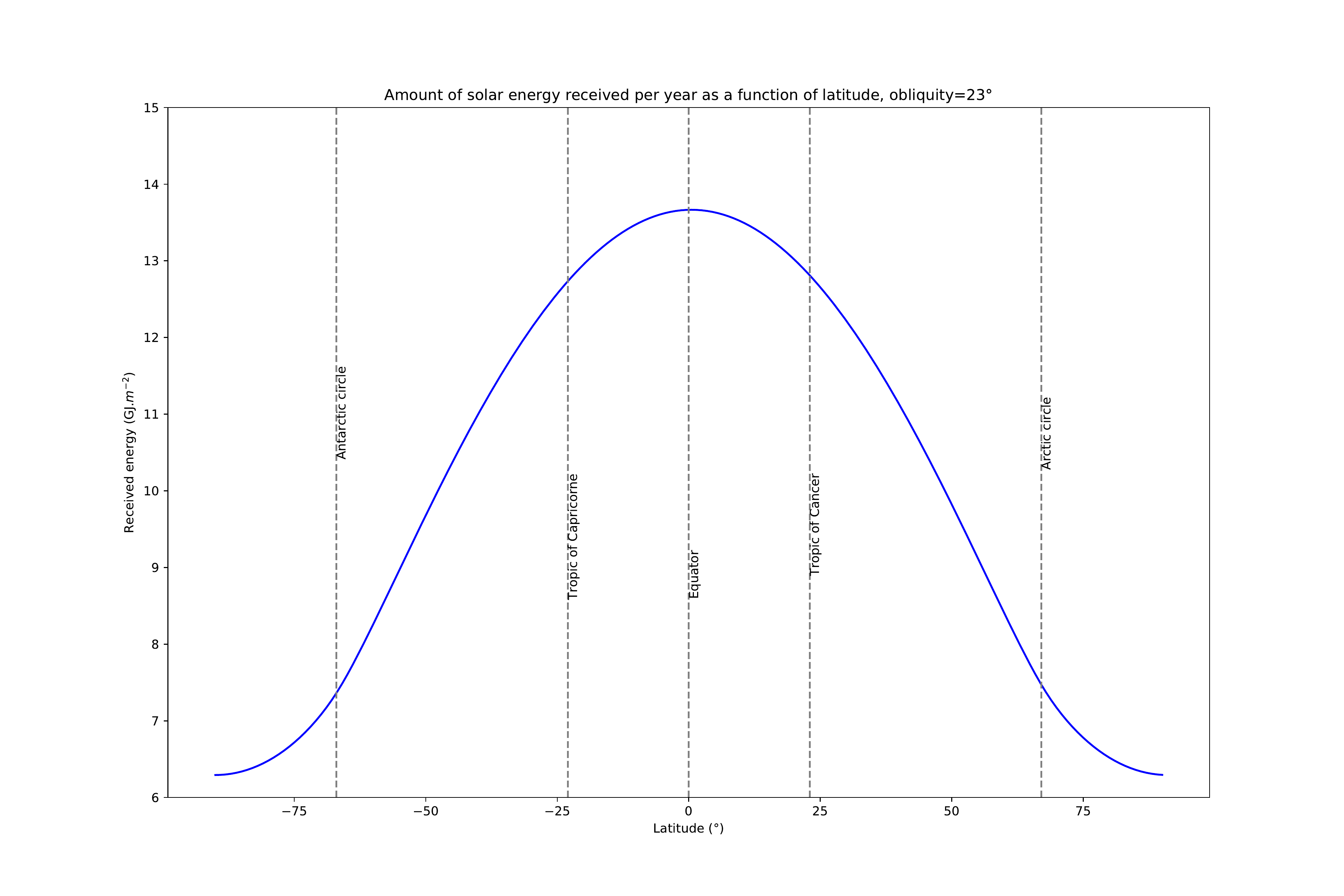}

\section{At which point can we observe sun at zenith ?}\label{s.zenith}
We are looking for which values of
$\lat $ it is
 possible to observe a day $ \jr $ in the year at
 an hour $ \hr $ the sun at its zenith,
 and if this phenomenon is possible in this latitude, for which values $ \jr $ and $ \hr $.
First, we reason without taking into account the phase shift.
The sun is at its zenith on day $ t $ at universal time (i.e. without the phase shift) $ \hr $ at latitude $ \lat $ if and only if
%
\[
\begin{pmatrix}
x_{\lat,\obl,\hr,t}  \\
y_{\lat,\obl,\hr,t}  \\
z_{\lat,\obl,\hr,t}
\end{pmatrix} 
= \begin{pmatrix}
0 \\
1  \\
0
\end{pmatrix} 
\]
i.e. if and only if
\[
\begin{cases}
\cos(\obl)\cos(\lat)\cos(\jr)\cos(\hr)-\sin(\obl)\sin(\lat)\cos(\jr)+\cos(\lat)\sin(\jr)\sin(\hr)
=0\\
-\cos(\obl)\cos(\lat)\sin(\jr)\cos(\hr)-\sin(\obl)\sin(\lat)\sin(\jr)+ \cos(\lat)\cos(\jr)\sin(\hr)    = 1
\\                                                                                         
\sin(\obl)\cos(\lat) \cos(\hr)  +  \cos(\obl) \sin(\lat)    =0.
\end{cases}
\]
Thanks to last line, we can replace $\cos(\hr)$ by $\frac{-\sin(\lat) \cos(\obl)}{\cos(\lat)\sin(\obl)}=-\frac{\tan(\lat)}{\tan(\obl)}$.
In addition, by operating a phase shift of $ \delta $, we noticed that the maximum, at latitude and fixed day (ie $ \obl, \lat, \jr $ fixed) of $ p _{\obl} (\lat, t, \hr) $ is reached in $ \hr = 0 $ so if you want to hope to see the sun at its zenith, you must be at the time offset $ 0 $, so at universal time $ - \delta $, but we know from the above \eqref {def.delta} that
this number checks

Since cosinus is an even function, we have $\cos(-\delta)=\frac{\cos(\obl)\sin(\jr)}{\sqrt{\left(\cos(\obl)\sin(\jr)\right)^2+\cos^2(\jr)}}$from  what
\begin{equation}
\cos(\delta)=\frac{a}{a^2+b^2} =\frac{\cos(\obl)\cos(\jr) }{\sqrt{\cos^2(\obl)\cos^2(\jr)+\sin^2(\jr)}}              
\end{equation} 
By squaring the two expressions we just identified at $ \cos (\delta) $, we get
\begin{equation*}
\frac{\tan^2(\lat)}{\tan^2(\obl)} = \frac{\cos^2(\obl)\cos^2(\jr)}{\left(\cos(\obl)\cos(\jr)\right)^2+\sin^2(\jr)}.
\end{equation*}
that is
\begin{equation*}
{\tan^2(\lat)}\left[{\left(\cos(\obl)\sin(\jr)\right)^2+\cos^2(\jr)}\right] = {\tan^2(\obl)}{\cos^2(\obl)\sin^2(\jr)}={\sin^2(\obl)}{\sin^2(\jr)}.
\end{equation*}
Then, using $\cos^2=1-\sin^2$
\begin{equation*}
{\tan^2(\lat)}\left[\left(\cos^2(\obl)\sin^2(\jr)\right)^2+1-\sin^2(\jr)\right] = \sin^2(\obl)\sin^2(\jr).
\end{equation*}
and simplifying
\begin{equation*}
\tan^2(\lat)\left(\cos^2(\obl)-1\right)\sin^2(\jr)+\tan^2(\lat) = \sin^2(\obl)\sin^2(\jr).
\end{equation*}
we use again  $\cos^2-1=-\sin^2$ :
\begin{equation*}
-\sin^2(\obl)\tan^2(\lat)\sin^2(\jr)+\tan^2(\lat) = \sin^2(\obl)\sin^2(\jr).
\end{equation*}
Gathering terms appearing as factor of $\sin^2(\jr)$ :
\begin{equation*}
\tan^2(\lat) = \left(\sin^2(\obl) +\tan^2(\lat)\sin^2(\obl)\right)\sin^2(\jr).
\end{equation*}	
Using $1+\tan^2= \frac{\cos^2+\sin^2}{\cos^2}=\frac{1}{\cos^2}$ :
\begin{equation*}
\tan^2(\lat) =\frac{\sin^2(\obl)}{\cos^2(\lat)} \sin^2(\jr).
\end{equation*}
We find
\begin{equation*}
\sin(\jr)= \pm \frac{\sin(\lat)}{\sin(\obl)}.
\end{equation*}
For there to be values of $ \jr $ solution of this equation, it is necessary that the value $ \frac{\sin(\lat)}{\sin (\obl)} $ is realizable by the function sine, in other words that it belongs to its image which is $ [- 1,1] $. This only occurs for latitudes $ \lat $ satisfying $ | \sin (\lat) | \le \sin (\obl) $, i.e. $ \lat \in [- \obl, + \obl] $.
which correspond to the points between the two tropics. For $ \lat = 0 $, the equator, the moments of the year $ \jr $ at which this phenomenon occurs are the equinoxes: $ \jr = 0 $ or $ \pi $. For points in the tropics, $ \lat = \obl $ (resp. $ - \obl $), this occurs when $ \sin (\jr) = 1 $, ie when $ t = \frac {\pi}{2 } $: summer solstice (resp. $ t = \frac {3 \pi}{2} $: southern summer solstice). Between the tropics, this occurs at noon (out of phase) on two days of the year:
\begin{equation}
t_{zenith}({\lat,\obl}) := \arcsin\left( \frac{\sin(\lat)}{\sin(\obl)} \right) \text{ ou } \pi - \arcsin\left(\frac{\sin(\lat)}{\sin(\obl)}\right).
\end{equation}

\section{Direction of sunsets}\label{s.lever_coucher_soleil}
For the anecdote, arriving in Reunion, I often wandered around the barachois. I really liked going to see the sun go down there, to the west overall, slightly to the north since we are in the southern hemisphere ($ 21 ^ \circ $ south). All this made him lie down in the sea, slightly to the right of a cliff below which the new coastal road was being built. And then one fine day, the sun began to set behind the Mountain (capital because it is the name of a town of St Denis). Having already started well in the writing of this file, I told myself that I could solve the kind of question: at such latitude, from what date to what date the sun will set in this or that direction.

Before considering any rotation, let's put the globe with the south pole north pole axis vertically, and plot at the point of latitude $ \lat $ and longitude $ 0 $ the two vectors that generate the plane tangent to the sphere at this point. We call these vectors $ \overrightarrow {e} _N $ and $ \overrightarrow {e} _O $ to signify that the first is the unit vector in the north direction and the second in the west direction. We then have
%
%
\[
\overrightarrow{e}_o=\begin{pmatrix}
0\\-1\\0
\end{pmatrix}
\text{ et }
\overrightarrow{e}_n=\begin{pmatrix}
\sin(\lat)\\0\\\cos(\lat)
\end{pmatrix}.
\]
Let's take again the notations of the section \ref{s.notations}
After rotation of an angle $ \hr $ around the axis of $ z $ then
According to \eqref {def.delta}, the phase shift that we had achieved was worth
\[\delta = \arctan\left(
\frac{-\tan(\jr)}{\cos(\obl)}\right) \quad(+\pi \text{ si }\jr\in\left[\frac{\pi}{2},3\frac \pi 2\right]).\]
This phase shift was convenient to center the day at noon but here, it is the real value of $ \hr $ that will matter to us. With phase shift, always according to \eqref {e.def.a_obl}, we have
%
%
so without phase shift, the respective angles of sunrise and sunset are
\begin{equation}\label{e.edf.debutjournee}
\hr_{lever}=	\arctan\left(
\frac{\tan(\jr)}{\cos(\obl)}\right)
- 
\arccos\left(\frac{ \sin(\obl)\sin(\lat)\cos(\jr)}{\cos(\lat) \sqrt{\cos^2(\obl)\cos^2(\jr)+\sin^2(\jr)}}\right)
\end{equation}
and 
\begin{equation}\label{e.edf.finjournee}
\hr_{coucher}=\arctan\left(
\frac{\tan(\jr)}{\cos(\obl)}\right)
+
\arccos\left(\frac{ \sin(\obl)\sin(\lat)\cos(\jr)}{\cos(\lat) \sqrt{\cos^2(\obl)\cos^2(\jr)+\sin^2(\jr)}}\right).
\end{equation}
Remembering that $ \cos (a + b) = \cos (a) \cos (b) - \sin (a) \sin (b) $ (because real part of a product is worth product of real parts minus product imaginary parts), as well as $ \cos (\arccos (x)) = x $, $ \sin (\arccos (x)) = \pm \sqrt {1-x ^ 2} $ and $ \cos (\arctan (x)) = \pm \frac {1}{1 + x ^ 2} $, and $ \sin (\arctan (x)) = \pm \frac {x}{1 + x ^ 2} $ with the sign to be determined with a little common sense, we find, by putting $ u $ and $ v $ the respective arguments of the arctangent and the arccosine:

\begin{eqnarray*}
	\cos(\hr_{sunset})&=&\cos(\arctan(u))\cos(\arccos(v))-\sin(\arctan(u))\sin(\arccos(v))\\
	&=&\frac{v-u\sqrt{1-v^2}}{1+u^2}
\end{eqnarray*}
et
\begin{eqnarray*}
	\sin(\hr_{couset})&=&\cos(\arctan(u))\sin(\arccos(v))+\sin(\arctan(u))\cos(\arccos(v))\\
	&=&\frac{u\sqrt{1-v^2}+v}{1+u^2}.
\end{eqnarray*}

If we consider the sun as a point at infinity in the direction $ Ox $, the vector between the point where we are and the sun is always
$e_x=\begin{pmatrix}
1\\0\\0
\end{pmatrix}$,
a fortiori at the time of sunset. To have the components in the local coordinate system
 $(\overrightarrow{e}_N,\overrightarrow{e}_O)$, 
 it is `` enough '' to project this vector on the two elements of the base:
  $(\bigmatrice_{\obl,\jr,\lat,\hr_{coucher}} e_y,\bigmatrice_{\obl,\jr,\lat,\hr_{coucher}} e_z)$. 
  The vector $ e_x $ actually belongs to the plane generated by these two vectors, because the moment of sunset precisely corresponds to the moment when the next component
$\bigmatrice_{\obl,\jr,\lat,\hr_{coucher}} e_x$ equals zero. 
\begin{itemize}
	\item And the western component is worth $\bigmatrice_{\obl,\jr,\lat,\hr_{coucher}} e_y\cdot e_x = (\bigmatrice_{\obl,\jr,\lat,\hr_{coucher}})_{1,2}$ i.e.
	\[
	\sin(\jr) \cos(\hr_{coucher})-\cos(\obl) \cos(\jr) \sin(\hr_{coucher})\]
	\item And the northern component is worth $\bigmatrice_{\obl,\jr,\lat,\hr_{coucher}} e_z\cdot e_x = (\bigmatrice_{\obl,\jr,\lat,\hr_{coucher}})_{1,3}$ i.e.
	\[
	\cos(\jr) (-\cos(\obl) \sin(\lat) \cos(\hr_{coucher})-\sin(\obl) \cos(\lat))-\sin(\jr) \sin(\lat) \sin(\hr_{coucher})
	\]
\end{itemize}
Actually, the data of only one of the two preceding quantities makes it possible to deduce the other (except for the sign, which can then be determined according to the season)
because the sum of their squares is $ 1 $, the vector $ e_x $ being unitary. If we are interested only in the direction, ie by the angle formed between $ e_x $ and the direction EAST, we know that its cosine is worth the opposite of the western component, its sine being worth the northern component (thus the sign of the sine allows to know if the angle is worth $ +\arccos (...) $
 or $ - \arccos (...) $), and its tangent the quotient north component/component west. 
 The reader can develop the calculation by replacing the values of
 the cosines and sines of $ \hr_{sunset} $ (resp. $ \hr_{raise} $) by the values 
 obtained previously as a function of $ u $ and $ v $, then to find out if there are simplifications. Numerically, we obtain this (in polar night situation, we represent the angle as being worth $ 90^\circ $ and in polar day situation as being worth $ -90^\circ $: in these cases, there at sunset or sunrise, and this convention allows to represent continuous functions):

\hspace{-3cm} \includegraphics[scale=0.5]{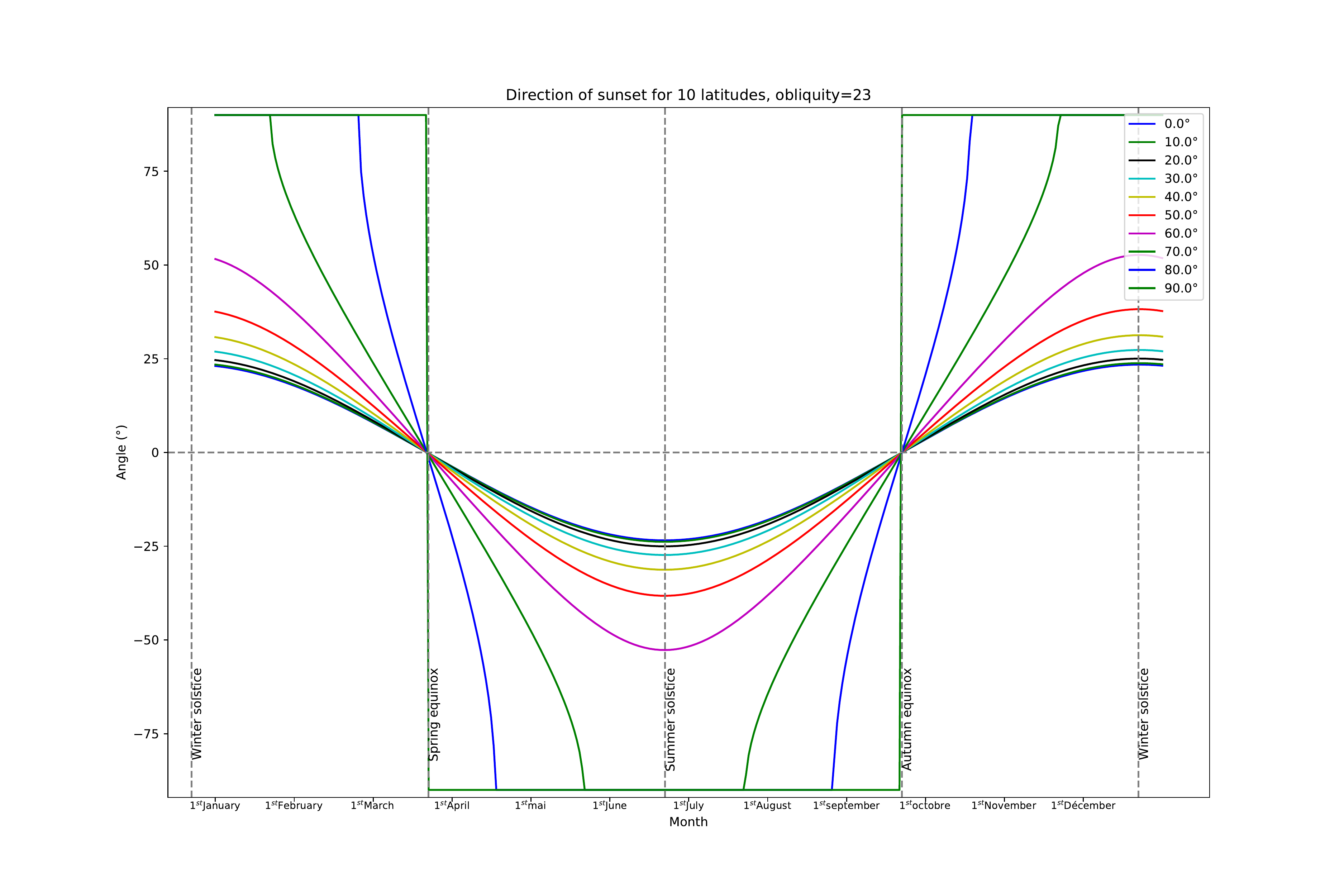}

\bibliographystyle{plain}
\bibliography{sun}

\end{document}